\documentclass[twocolumn,showpacs, aps, superscriptaddress, eqsecnum, prd, 
notitlepage, showkeys, nofootinbib, floatfix]{revtex4-2}

\usepackage{graphicx}
\usepackage{dcolumn}
\usepackage{bm}

\usepackage{amssymb}
\usepackage{amsmath}
\usepackage{graphicx}
\usepackage{dcolumn}
\usepackage{hyperref}
\usepackage{color,units}
\usepackage[dvipsnames]{xcolor} 
\usepackage{aas_macros}
\usepackage{lineno}
\usepackage{xspace}
\usepackage[normalem]{ulem} 
\usepackage{float}
\usepackage{orcidlink}
\usepackage{tabularray}

\newcommand{\CornellPhysics}{\affiliation{Department of Physics, Cornell University, Ithaca, NY, 14853, USA}}
\newcommand{\Cornell}{\affiliation{Cornell Center for Astrophysics and Planetary Science, Cornell University, Ithaca, New York 14853, USA}}
\newcommand{\Caltech}{\affiliation{Theoretical Astrophysics, Walter Burke Institute for Theoretical Physics, California Institute of Technology, Pasadena, California 91125, USA}}

\hypersetup{colorlinks=true, citecolor=teal, urlcolor=teal, linkcolor=teal}

\usepackage[commentmarkup=uwave,commandnameprefix=ifneeded]{changes}
\definechangesauthor[name=Max, color=teal]{MI}

\begin{document}

\title{Striking the right tone: toward a self-consistent framework for measuring black hole ringdowns}

\author{Teagan A. Clarke \orcidlink{0000-0002-6714-5429}}
\email{teagan.clarke@monash.edu}

\affiliation{School of Physics and Astronomy, Monash University, VIC 3800, Australia}
\affiliation{OzGrav: The ARC Centre of Excellence for Gravitational Wave Discovery, Clayton, VIC 3800, Australia}

\author{Maximiliano Isi \orcidlink{0000-0001-8830-8672}}
\affiliation{Center for Computational Astrophysics, Flatiron Institute, New York NY 10010, USA}
\author{Paul D. Lasky \orcidlink{0000-0003-3763-1386}}
\affiliation{School of Physics and Astronomy, Monash University, VIC 3800, Australia}
\affiliation{OzGrav: The ARC Centre of Excellence for Gravitational Wave Discovery, Clayton, VIC 3800, Australia}
\author{Eric Thrane \orcidlink{0000-0002-4418-3895}}
\affiliation{School of Physics and Astronomy, Monash University, VIC 3800, Australia}
\affiliation{OzGrav: The ARC Centre of Excellence for Gravitational Wave Discovery, Clayton, VIC 3800, Australia}

\author{\\ Michael Boyle \orcidlink{0000-0002-5075-5116}} \Cornell
\author{Nils Deppe \orcidlink{0000-0003-4557-4115}} \Cornell\CornellPhysics
\author{Lawrence E.~Kidder \orcidlink{0000-0001-5392-7342}} \Cornell
\author{Keefe Mitman \orcidlink{0000-0003-0276-3856}}\Caltech
\author{Jordan Moxon \orcidlink{0000-0001-9891-8677}} \Caltech
\author{Kyle C. Nelli \orcidlink{0000-0003-2426-8768}} \Caltech
\author{William Throwe \orcidlink{0000-0001-5059-4378}} \Cornell 
\author{Nils L. Vu \orcidlink{0000-0002-5767-3949}} \Caltech

\begin{abstract}
The ringdown portion of a binary black hole merger consists of a sum of modes, each containing an infinite number of tones that are exponentially damped sinusoids. 
In principle, these can be measured as gravitational-waves with observatories like LIGO/Virgo/KAGRA, however in practice
it is unclear how many tones can be meaningfully resolved. 
We investigate the consistency and resolvability of the overtones of the quadrupolar $\ell = m = 2$ mode 
by starting at late times when the gravitational waveform is expected to be well-approximated by the $\ell m n = 220$ tone alone. 
We present a Bayesian inference framework to measure the tones in numerical relativity data. 
We measure tones at different start times, checking for consistency: we classify a tone as stably recovered if and only if the 95\% credible intervals for amplitude and phase at time $t$ overlap with the credible intervals at all subsequent times.
We test a set of tones including the first four overtones of the fundamental mode and the 320 tone and find that the  220 and 221 tones can be measured consistently with the inclusion of additional overtones. 
The 222 tone measurements can be stabilised when we include the 223 tone, but only in a narrow time window, after which it is too weak to measure. 
The 223 tone recovery appears to be unstable, and does not become stable with the introduction of the 224 tone. 
We find that $N=3$ tones can be stably recovered simultaneously.
However, when analysing $N \geq 4$ tones, the amplitude of one tone is consistent with zero.
Thus, within our framework, one can identify only $N=3$ tones with non-zero amplitude that are simultaneously stable.

\end{abstract}

\maketitle

\section{Introduction}
The final stage of a binary black hole coalescence, called the ringdown, consists of a perturbed remnant black hole emitting gravitational waves. 
In general relativity, the gravitational waves from the ringdown can be decomposed using spin-weighted spheroidal harmonics into quasi-normal modes \citep[QNMs, see][]{Vishveshwara_1970, Teukolsky_1973, Chandrasekhar_1975, Leaver_1985, Kokkotas_1999}.
Each spheroidal harmonic mode is labeled with indices $\ell \geq 2$ and $|m| \leq \ell$.
There are an infinite number of tones associated with each angular mode, each denoted with $n \geq 0$.
The frequency and damping time of each tone depends only on the mass and spin of the remnant black hole (assuming zero charge) according to the no-hair theorem \citep[e.g.,][]{Carter_1971}. 

Understanding how to measure black hole tones can allow us to undertake tests of general relativity and the no-hair theorem for black holes \cite{Dreyer2004,Berti_2006,Isi_2019,LIGOScientific:2020tif,Cotesta_2022,LIGOScientific:2021sio}. 
However, the start of the ringdown---defined as the time when the signal can be described with black hole perturbation theory---is ambiguous \citep[e.g.,][]{Nollert_1996, Thrane_2017, Bhagwat_2018, isi_2021_analysing}. 
Understanding how early the perturbative prescription can be applied to the signal is key to correctly performing tests of general relativity and the no-hair theorem. 
Beginning the analysis too early could result in over-fitting to non-linear features in the signal \citep[e.g.,][]{TGR_2016, Bhagwat_2020, Baibhav_2023}.
If one waits too long to begin the analysis, the strain amplitude will have decayed exponentially, making spectroscopic tests impossible given the finite sensitivity of our instruments~\cite{Thrane_2017, Bustillo_2021}.
While black hole spectroscopy is part of a broader effort to test general relativity with gravitational waves, along with e.g., inspiral-merger-ringdown tests \citep[e.g.,][]{Dreyer2004, Hughes2005, Ghosh_2017, Johnson-McDaniel2022}, the ringdown provides one of the most direct ways to test the no-hair theorem. 
Binary parameter-informed analyses of the ringdown provide an alternative approach to damped sinusoid models for QNM extraction and tests of the no-hair theorem \citep[e.g.,][]{Brito2018, Gennari2023}.

Numerical relativity simulations provide numerical solutions to the field equations \citep{Pretorius_2005, Campanelli_2006, Baker_2006}, and are the state of the art for investigating the QNM decomposition of the ringdown. However, the components contained in these solutions remain unclear. Black hole perturbation theory provides predictions for the frequency and damping times of each tone, but the optimal time to fit them and how many tones can be meaningfully extracted remains unclear due to, e.g., non-linearities, non-orthogonal QNM decomposition, and error in the simulations themselves. Much effort has been devoted to determining how many tones measured at what time provide the optimal fit of physically-motivated tones to the numerical simulations. 

Many studies \citep[e.g.,][]{Buonanno_2007, Baibhav_2018, Ota2020, Mourier2021, Sago2021, Li_2022} find that several overtones of the $\ell m = 22$ mode can be extracted at early times after the merger and that they are needed to infer the correct mass and spin of the black hole. 
Ref.~\cite{Giesler_19} suggests that the linear QNM model may be valid as early as the strain peak time.
They show that higher overtones of the 22 mode up to $n = 7$ improve fits to simulated gravitational-waveforms from the strain peak time. Ref.~\cite{Ma_2022} fit away these seven tones from the NR data using frequency domain filters, uncovering evidence for second-order effects and spherical-spheroidal mode mixing.

Refs.~\cite{Cook2020, Zertuche_2022} find that seven tones of the $\ell=m=2$ as well as many tones from additional QNMs can be included to improve the fits. Tones up to $n = 9$ have also been explored for their potential to further improve fits at the strain peak time \citep{Forteza_2021}.
The apparent linearity of the signal at such early times could be explained if the non-linear effects are hidden behind the apparent horizon of the black hole \citep[e.g.,][]{Okounkova_2020, Chen_2022}, unable to reach the observer at infinity. On the other hand, non-linear effects like second order QNMs, may be required to accurately model the ringdown of higher harmonics \citep[e.g.,][]{Mitman_2023, Cheung_2023a}, although the magnitude of such second-order contributions are by far subdominant for existing detections.

While it seems clear that the early ringdown can be modeled with a large number of overtones, there is debate as to whether the associated fits are physical and to which extent \citep[e.g.,][]{Pook-Kolb2020, Mourier2021},
i.e., are we overfitting tones to produce what amounts to a phenomenological model?

Refs.~\citep{London_2014, London_2020} construct analytic fits for stable tones as a function of binary parameters. They point out that in the purely perturbative QNM regime, tone amplitudes and phases should be measured consistently at different start times. Ref.~\cite{London_2020} find that the amplitudes for tones with $n>0$ are not consistent in time and omit these from their analytic fits. However, \cite{JimenezForteza2020} finds that analytic fits for tones are biased when fitting at early times, possibly indicating the presence of higher tones and non-linearities near the strain peak. 

Ref.~\cite{Baibhav_2023} further explore whether or not the ringdown tones are consistent when measured at different start times.
They argue: if the overtone parameters do not yield consistent fits at different start times, they are not physical, but rather show evidence of overfitting.
Ref.~\cite{Zhu_2023} finds that overtones beyond 223 are always inconsistent by this metric when fitting in the time domain, while \cite{Nee_2023} use analytic modelling to caution that increasing the number of tones may end up overfitting to a misspecified model. Ref.~\cite{Baibhav_2023} similarly finds that adding tones improves the fits, even to unphysical hybrid waveforms,  which they point out as a sign of overfitting.
Ref.~\cite{Cheung_2023} attempts to resolve this issue by measuring the stability of the tones as they are being fitted, iteratively removing inconsistent tones until only stably recovered tones are left. 
For the 22 extraction of a numerical-relativity waveform, they find that five tones are stably recovered, with 221 being the highest stably measured overtone. The other robust tones consist of retrograde modes or modes with the same $m$ but different $\ell$, due to other angular QNMs mixing into the 22 spherical harmonic.
Reference.~\cite{Redondo-Yuste2023b} uses the Bayesian evidence to algorithmically select tones to be included in their model, and to search for non-linear contributions. They find that non-linear contributions may be important for ringdown analysis, although the importance decreases as the black hole spin increases. 

In parallel to the debate about the physicality of the linear perturbation fits, a number of studies have attempted to measure black hole overtones in  gravitational-wave data with mixed results. Assuming a QNM decomposition, Refs.~\cite{TGR_2016, Carullo_2019} search for overtones in the late-time ringdown of GW150914, only finding evidence of the fundamental 220 tone.  
 By studying at earlier times in the ringdown, Refs.~\cite{Isi_2019, isi_2021_analysing, Wang_2023b, Ma_2023} find evidence for the 221 tone in GW150914. Refs.~\cite{Finch_2022, Wang_2023, Crisostomi2023} find only weak support for the 221 tone in GW150914. However, \cite{Cotesta_2022, Correia_2023} do not find evidence for any higher tones beyond the fundamental in GW150914 (although see Refs.~\cite{Isi:2023nif,Carullo:2023gtf} for further discussion on these results).

In this paper, we seek to arrive at a \textit{self-consistent}, perturbative model of the post-merger ringdown signal.
We present a Bayesian-inference, forward-modelling procedure as a method for extracting the ringdown QNMs from numerical-relativity simulations. Using a numerical-relativity waveform, we start at the end of the ringdown when the strain should be dominated by the 220 (and 320) tones and use this to fit the $n=0$ amplitude and phase.

We also fit a noise amplitude---a phenomenological tool, which we introduce in order to account for various unmodeled physics---to cover anything that causes the late-time ringdown to deviate from a pure 220 tone, apart from some small mode-mixing contribution from the 320 tone.
Having established the asymptotic behaviour of the ringdown signal, we work backward, carrying out Bayesian inference to see if there is support for additional tones at earlier times.

We assess whether each tone in a set is stably recovered by checking whether the fits obtained from different times are consistent.
In doing so, we aim to determine if a numerical-relativity waveform can be self-consistently modeled using a superposition of tones, and how many tones can be said to be present.

The remainder of this paper is organised as follows.
In Section~\ref{sec:framework} we describe our model and fitting procedure. 
In Section \ref{sec:results} we present our results before discussing their implications in Section \ref{sec:discussion}.

\section{Framework}\label{sec:framework}

We model the ringdown strain as a sum of damped sinusoids, writing each spin-weighted spherical harmonic mode of the strain as a complex-valued time-series: $h = h_+ - ih_\times$ as
 \begin{equation}
     h_{\ell m}^{N_\text{tones}} (t) = \sum^{N_\text{tones}}_{n=0}A_{\ell m n}e^{-i[\omega_{\ell m n}(t - t_0) + \phi_{\ell m n}]},
\label{eq:model}
\end{equation}
where $\omega_{\ell m n}(M_f,\chi_f)$ are the complex frequencies determined by the remnant mass and spin through the no-hair theorem \citep{Berti_2009, berti_website}. 
The negative imaginary component of $\omega_{\ell m n }$ is the inverse tone damping time $\tau_{\ell m n}$. 
The tone amplitudes $A_{\ell m n}$ and phases, $\phi_{\ell m n}$ depend on the initial conditions of the perturbations on the black hole. 
The variable $t_0$ is the reference time of the fit, which we take as the peak time of the 22 mode strain, $t = 0M$. We use geometric units in this study and measure $t$ in units of the initial binary mass $M$, which is set to unity.

We fit the $\ell=m=2$ mode alone and vary $N_\text{tones}$ to change the number of tones in the fit. We order the tones in descending order of damping time: 220, 221, ..., $22N_\text{tones}$. 
We fit numerical relativity data that has been decomposed into spin-weighted spherical-harmonics $_{-2}Y_{\ell m}$.
However, the basis for a perturbed black hole is actually described by spin-weighted \textit{spheroidal} harmonics $_{-2}S_{\ell m}$ \citep[e.g.,][]{Teukolsky_1972, Teukolsky_1973}. 
Because of this, mode-mixing occurs between QNMs with the same $m$ but different $\ell$. 
For the fundamental 22 mode, the dominant source of mode-mixing comes from the 320 \citep[e.g.,][]{Buonanno_2007}, which has a higher frequency than the 220 and a comparable damping time. To account for this, we also include the 320 in our fits.

The data for our fit is from the Simulating eXtreme Spacetimes (SXS) catalogue \citep{SXS_2013,SXS_2019}.
We use the Cauchy characteristic evolved (CCE) waveform simulation \citep{Mitman_2020, Moxon_2020, Moxon_2021}, which was generated using the \texttt{SpECTRE} code’s CCE module \citep{SPECTRE}. The waveform is mapped to the superrest frame of the remnant black hole with Bondi-van der Burg-Metzner-Sachs (BMS) frame fixing \citep{Mitman_2021, Zertuche_2022, Mitman_2022}. This is the correct frame mapping for QNM extraction and eliminates unphysical frame effects. 
The frame fixing is performed using the \texttt{scri} python module \citep{Boyle_2013, Boyle_2014, Boyle_2016, scri}. We use the \texttt{SXS:0305} waveform, which corresponds to the most likely parameters for the GW150914 LIGO-Virgo observation \citep{GW150914}. The simulation produces a remnant mass $M_f = 0.952 M$ and dimensionless spin $\chi_f = 0.692$. 
We use the highest available resolution for this study, which has a time resolution of $\approx 0.1 M$ around the strain peak time. We do not interpolate to a uniform time grid for this study. 

We use the MCMC sampler \texttt{emcee} \citep{emcee_2013} to fit a damped sinusoid model according to Eq.~\eqref{eq:model}, fitting the tone amplitude and phase as a function of time. 
We fix the frequency and damping time of each tone as a function of the known asymptotic remnant mass and spin from the numerical relativity simulation waveform metadata. It has been pointed out \citep[e.g.,][]{Buonanno_2007, Pook-Kolb2020, Pook-Kolb2020b, Sberna2022, Redondo-Yuste2023} that the mass and spin of the black hole may still evolve at early times in the ringdown, potentially impacting QNM fits. We do not consider this effect in this study. 
We calculate the frequency and damping times using the \texttt{qnm} python package for Kerr black holes \citep{Stein_qnm_2019}. 
We parameterise the amplitude and phase as $x=A\text{cos}(\phi)$ and $y=A\text{sin}(\phi)$ and sample in normal distributions of $x$ and $y$ with mean 0 and standard deviation 1.

In order to carry out Bayesian inference, we must introduce some notion of uncertainty.
One path forward could be to set the noise at the level of estimated numerical-relativity precision.
However, this numerical noise does not capture the systematic uncertainty in our calculation.
Our premise is that the very end of the numerical-relativity waveform should be consistent with a pure 220 tone with small contributions from the 320 tone and late-time polynomial tails \citep[e.g.,][]{Leaver_1986, Carullo2023}, although the latter are yet to be seen in numerical simulations from the SXS catalogue.
We use this late-time fit as a point of reference for understanding the earlier part of the waveform.
If the end of the waveform is \textit{inconsistent} with a set of tones comprising of only the 220 and 320 tones, this additional structure represents a systematic error within our framework.

 \begin{figure}
    \centering
    \includegraphics{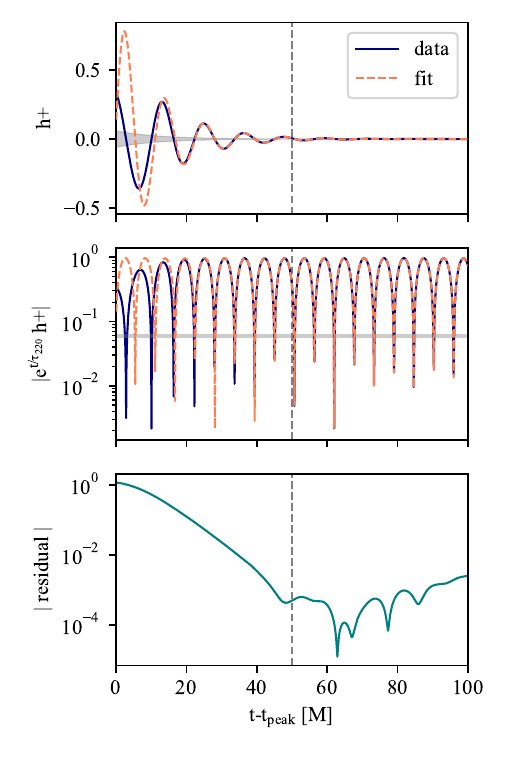}
    \caption{
    Ringdown time series.
    The top panel shows the 22 mode numerical-relativity data $h_+(t)$, which is the real part of the 22 mode strain, in solid blue alongside a fit obtained using just the 220 and 320 tones predicted by perturbation theory.
    The dashed vertical line represents the start time for this fit ($50M$). The grey band represents the artificial noise $\sigma_\text{sys}$. 
    The second panel shows the same timeseries, but multiplied by $e^{t/\tau_{220}}$ in order to counteract the exponential decay of the 220 tone. The horizontal line represents our model for the uncertainty $\sigma_\text{sys}$ in this regime, which in this representation is constant. 
    The bottom panel shows the difference between the numerical-relativity waveform and our fit as a function of time.    }
    \label{fig:timeseries}
\end{figure}

In order to quantify this systematic uncertainty, we fit the late waveform (starting from $t=100 M$), allowing for a mixture of the 220, 320 and 221 tones. We choose $100M$ as the earliest time when the 221 is beginning to become unresolved for an analytic waveform test consisting of three tones with the same frequency and damping times of the 220, 320 and 221 in the numerical-relativity simulation and amplitudes and phases consistent with our fits to the numerical-relativity data. 

Ideally one would wait longer than $100M$. However, the numerical-relativity error starts to increase from $100M$, suggesting the numerical relativity error starts to become more relevant from this point.\footnote{The numerical relativity error is the difference between the two highest resolution waveforms after they are mapped to the same BMS frame.}
We model the systematic uncertainty so that the \textit{artificial} noise on the strain is drawn from a Gaussian distribution with width $\sigma_\text{sys}$. The artificial noise is implemented by dividing the data by the exponential term of the 220 tone, $e^{-t/\tau_{220}}$, before performing our fits. $\tau_{220}$ is the damping time of the 220 tone $\approx 11.74M$, calculated using the simulation remnant mass and spin in the \texttt{qnm} package \citep{Stein_qnm_2019}.

We vary $\sigma_\text{sys}$ until the posterior for the amplitude of the 221 tone $A_{221}$ is consistent with zero at one-sigma credibility. 
This enforces our requirement that the late-time ringdown should not contain any contribution from overtones higher than the 220.
We obtain a value of $\sigma_\text{sys} = 0.06 $, which is $\approx 70$ times higher than the expected numerical-relativity error at $t=100 M$.
This suggests that our ability to understand the late-time behaviour of the ringdown is likely limited not by numerical-relativity noise, but by theoretical uncertainty about the relative contribution of various tones and/or non-linearities. Additional contributions from mode-mixing and power-law tails may also be included in our measurement of $\sigma_\text{sys}$.\footnote{Presumably, if one could obtain a sufficiently long and accurate numerical-relativity waveform, $\sigma_\text{sys}$ would approach the level of the numerical-relativity noise.}
The quantity $\sigma_\text{sys}$ has no physical meaning; it is a purely phenomenological tool that we introduce in order to quantify our present theoretical uncertainty about the behavior of the late-time ringdown. We find that the value of $\sigma_\text{sys}$ measured is not sensitive to the chosen time from approximately $t=70 M$ with $\sigma_\text{sys}$ changing by only $\approx 0.01$ between $70 M$ and $105 M$. The exact value of $\sigma_\text{sys}$ and the time at which it is measured is not important for our investigation into the stability of sets of tones. 
Our decision to model $\sigma_\text{sys}$ with an artificial uncertainty (rather than an absolute uncertainty) is motivated by experimentation. 
We find that our fits are relatively consistent in size and magnitude and that our sampling is better behaved when we assume an artificial uncertainty as a function of time, while they become difficult to compare assuming a fixed absolute uncertainty.
Our model for $\sigma_\text{sys}$ is shown in Fig.~\ref{fig:timeseries}. 
The top panel shows the real part of the 22 mode numerical-relativity waveform $h_+(t)$ and fit. We include a grey band representing our choice for the artificial noise, while the middle panel shows $h_+(t)$ multiplied by $e^{t/\tau_{220}}$; the artificial noise $\sigma_\text{sys}$ can be visualised as a horizontal line at 0.06 on this panel.
The bottom panel is a time series showing the difference between the numerical-relativity waveform and our fit.
We provide more details on this approach in Appendix \ref{app:noise}. 

\section{Results}
\label{sec:results}

\subsection{Consistency of tones with time}
We seek to find a set of tones that can produce a self-consistent measurement of the ringdown signal across time. 
To this end, it is useful to introduce a criterion to determine if each tone in a set is stably recovered.
For a tone to be recovered stably at time $t_0$, we require that the 95\% credible intervals for the amplitude and phase of that tone overlap with the 95\% credible intervals of \textit{all} subsequent fits (corresponding to larger values of $t_0$).
Thus, e.g., the fit at $10 M$ must be consistent with all the fits between $10 M$ and $70 M$ to be considered stable at $10 M$. 
We are interested in tones that are both consistent and resolved. In order for a tone to be considered stable and resolved at time $t_0$, we further require that the 95\% credible interval for the amplitude excludes zero.
This creates three possible classifications:
\begin{enumerate}
    \item \textbf{Stable recovery.} The amplitude of the tone is non-zero and the amplitude and phase of the tone are consistent with subsequent fits.
    \item \textbf{Unstable recovery.} The amplitude of the tone is non-zero, but the amplitude and/or phase are \textit{inconsistent} with subsequent fits.
    \item \textbf{Unresolved.} The amplitude of the tone is consistent with zero.
\end{enumerate}

\begin{figure}
    \centering
    \includegraphics{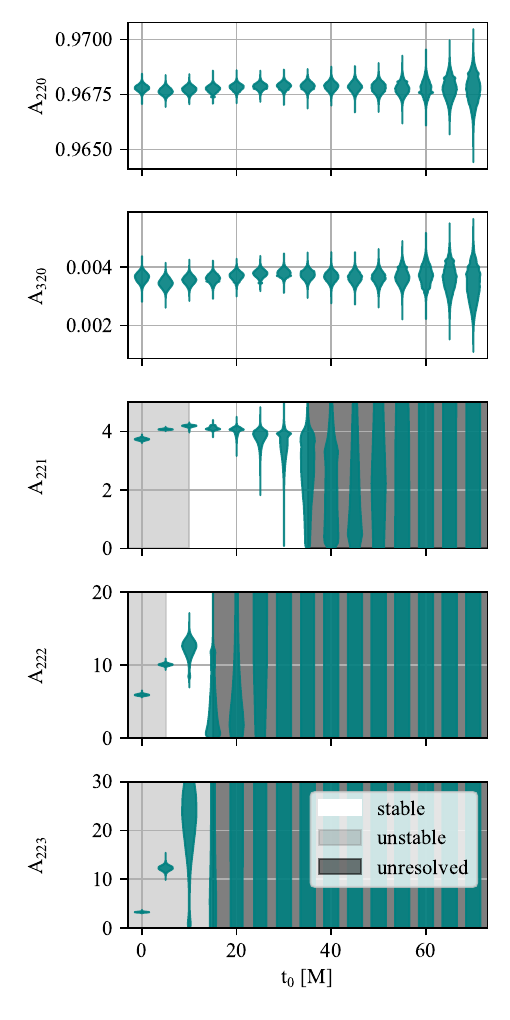}
    \caption{
    Posterior distributions for amplitude as a function of $t_0$. 
    Each panel represents a different tone.
    We show the regimes where each fit is stably recovered (white), unstably recovered (light gray) and unresolved (dark gray). We also fit the phase of each tone and require the phase to also be stable for that tone to be considered stable. For simplicity we only show the amplitude posteriors here.We show the corresponding phase posteriors in Fig.~\ref{fig:5phi_violin}.
    The 221, 222 and 223 tones are unstably recovered at early times and become unresolved at late times, which implies that they do not significantly improve the fit at late times. 
    The 223 tone is not stably recovered at any time but may stabilise with additional tones.}
    \label{fig:n5_fit}
\end{figure}

This classification method is illustrated in Fig.~\ref{fig:n5_fit}.
Each panel shows amplitude of a different tone as a function of start time $t_0$.
There are five tones in our fit: 220, 320, 221, 222 and 223. This amounts to four tones of the $\ell = m = 2$ QNM plus a contribution from the 320 due to mode-mixing.
At regularly spaced values of $t_0$, we plot the posterior for each amplitude in teal. The amplitudes are all measured at a reference time of $t = t_\text{peak}$ while the fit is performed at different start times $t_0$. 
The background is shaded according to the stability of the tone.
The light-gray regions indicate that the tone is unstably recovered, the white regions indicate it is stably recovered, and the dark-gray regions indicate that the tone is unresolved.
To avoid clutter, we do not show the accompanying plots of phase, although we also require phase consistency in order for a tone to be classified as stable. We provide the corresponding plot of the phase posteriors in Appendix~\ref{app:phase}. 

From the first two panels, we see that the 220 and 320 tones are stably recovered from $t_0 = 0M$. 
In Appendix~\ref{app:stability} we compare the 220 posteriors for the lowest and highest $n$ fits, demonstrating how the 220 stabilises with the inclusion of more tones. 
The 221 tone is initially unstably recovered, stabilises at $10 M$, and then becomes unresolved at $30M$.
The 222 tone is stably recovered within a narrow window around $5 - 15 M$ while the $223$ tone is unresolved.
With our framework, the $\ell = m = 2$ ringdown signal is well described by four tones which are stable from approximately $10 M$, when considering a model allowing 220, 320, 221, 222 and 223 contributions. 

We investigate how the stability changes with the number of tones in a set.
First we investigate the 220 alone. 
We find that the 220 is not stably recovered until approximately $30 M$ unless other tones are included in the fit.  
This is consistent with the expectation that the 220 should become dominant at late times. 
Adding the 320 does not improve the stability of the recovery significantly. 
When we add the 221, the 220 becomes stably recovered at $15 M$. 
The earliest stable time for the 220 changes to $5 M$ when we add the 222 and $0 M$ when we add the 223.\footnote{The fits in Section~\ref{sec:results} are all carried out assuming the frequency and damping times for each tone predicted by perturbation theory using the remnant mass and spin of the simulation.
In Appendix~\ref{app:frequency}, we provide the results of investigations where we treat the tone frequency as a free parameter.}

This analysis suggests that the 22 mode ringdown signal is well described by five tones. 
Adding the 224 tone may slightly improve the recovery stability of the 222, but does not improve that of the 223 and decays too quickly to be stably recovered itself. The 221 behaviour remains unchanged with the inclusion of the 224.
Additional tones produce fits consistent with zero amplitude.
Table~\ref{tab:stability} summarises the recovery stability of different tones as we vary the number of tones $N$.

We test the impact of reducing the spacing in start times $t_0$ on our results. 
We test the set of five-tones fit using a spacing of $2 M$ between $0$ and $20 M$ and find the tones behave consistently with the more coarsely spaced test. However, the 221 recovery is slightly improved, stabilising at $6 M$, rather than $10 M$. The 222 recovery stabilises at $8 M$--- slightly later than the previous test--- and becomes unresolved at $14 M$. The 223 recovery remains unstable at all times and becomes unresolved at $12 M$. We emphasise that the qualitative results are not significantly changed by the spacing of time steps.

We investigate the contribution of other potential sources of mode-mixing: the counter-rotating 220 tone and the 321 and 420 tones which were found to be visible in Refs.~\cite{Redondo-Yuste2023b, Cheung_2023}. 
We test the counter-rotating 220 (r220) in a set of tones with the 220 and 320 and find that the r220 tone is resolved only until 15M and its amplitude is smaller than the 320 ($< 0.003$). 
We test a set of tones including the five tones measured in Fig~\ref{fig:n5_fit} and the 321 tone. The 321 can be resolved until $20M$ with an amplitude of $\approx 0.1$ in the stable region of 5-15 M. The qualitative results of the other tones in this set remain unchanged, although the posteriors are in general broadened due to increased statistical uncertainty from introducing another parameter. We find that the 420 amplitude is consistent with 0 from $t=0M$ in a set of four tones (220, 320, 420, 221).

\begin{table}
  \centering
    \caption{
    Summary of when each tone is stably recovered for different values of $N_\text{tones}$ in a given set.
    The variable $t_\text{stable}$ indicates the time at which the amplitude and phase of each tone become consistent with subsequent fits using later parts of the waveform (and which the amplitude of the tone is inconsistent with zero).
    This time (and all other times in this table are quoted in units of $M$.
    If the tone is not stably recovered for any time, this cell is marked ``$-$''.
    The variable $t_\text{unresolved}$ indicates the time at which the amplitude becomes consistent with zero.
    If the tone is never consistent with zero, this cell is marked ``$-$''. $\tau$ denotes the damping time for each tone. 
    }

\renewcommand{\arraystretch}{0.5} 
\SetTblrInner{rowsep=0.8pt, colsep=5pt}
    \begin{tblr}{c c c c c c}
    \hline
    \hline
    \rule{0pt}{3ex}  
    N$_\text{tones}$ & tone & $\tau$ & t$_\text{stable} $ & t$_\text{unresolved}$ & stable interval \\
    \hline
        1 & 220 & 11.74 & 30 & $-$ &  $30-70$ \\
    \hline
        \SetCell[r=2]{} 2 & 220 & 11.74 & 30 & $-$ & \SetCell[r=2]{} $30-70$\\ 
          & 320  & 11.27 &  0 & $-$ \\
    \hline
     \SetCell[r=3]{} 3 & 220 & 11.74 & 15 & $-$ &  \SetCell[r=3]{} $30-45$\\
         & 320  & 11.27 &  5 & $-$ & \\
         & 221  & 3.88 & 30  & 45 & \\
    \hline
       \SetCell[r=4]{} 4 & 220 & 11.74 & 5 & $-$  & \SetCell[r=4]{}  $-$  \\
         & 320  & 11.27 & 0  & $-$ \\
         & 221  & 3.88 & 20 & 40 \\
         & 222  & 2.30 & $-$  & 20 \\
    \hline
     \SetCell[r=5]{} 5 & 220 & 11.74 & 0 & $-$ & \SetCell[r=5]{} $-$ \\
          & 320 & 11.27 & 0 & $-$ \\
          & 221 & 3.88 & 10 & 35 \\
          & 222 & 2.30 & 5 & 15 \\
          & 223 & 1.62 & $-$ & 15 \\
    \hline
    \SetCell[r=6]{} 6 & 220 & 11.74 & 0 & $-$  & \SetCell[r=6]{}  $-$ \\
      & 320 & 11.27 & 0 & $-$ \\
      & 221 & 3.88 & 10 & 35 \\
      & 222 & 2.30 & 0 & 15 \\
      & 223 & 1.62 & $-$ & 10 \\
      & 224 & 1.26 & $-$ & 5 \\
    \hline
    \hline 
    \end{tblr}

    \label{tab:stability}
\end{table}

\begin{figure}
    \centering
    \includegraphics{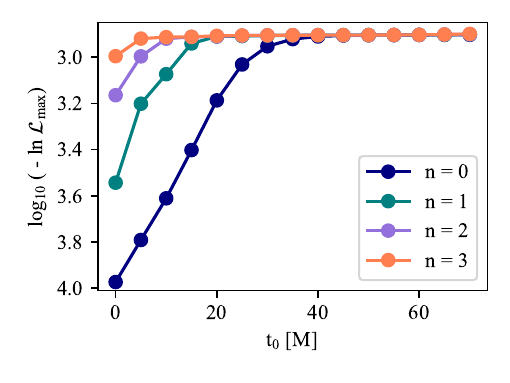}
    \caption{
    Maximum natural log likelihood as a function of $t_0$ for different values of $N_\text{tones}$. 
    (To handle the large dynamic range of likelihood values, we actually plot the $\text{log}_{10}$ of $-\ln {\cal L}_\text{max}$.)
    All four models include the 320 tone. 
    At early times the fits improve as we add tones to the model before saturating at around $30 M$, lending further support to to the conclusion that the fit is dominated by the 220 past $30 M$.}
    \label{fig:max_ln_L}
\end{figure}

As a consistency check, we assess the goodness-of-fit for each of our models by comparing the maximum log likelihood of each model as a function of time. Figure \ref{fig:max_ln_L} shows the log likelihood of each model as a function of start time $t_0$. 
The likelihood increases as the number of tones is increased, suggesting that the addition of tones serves to improve the fit, although the improvements are minimal from $30-40 M$ after the strain peak.

\subsection{How many tones can be resolved at the waveform peak?}

Ref.~\cite{Giesler_19} suggests that the perturbative regime can be applied as early as the waveform peak ($t_0 = 0 M$), and that seven tones produce the best fit at this time.  
We investigate the number of tones that can be resolved at $0 M$, when the strain amplitude is maximal. 
We fit our numerical-relativity data with a set of eight tones, consisting of the first seven tones of the 22 mode and the 320 tone. 
We find that overtones higher than 224 are not resolved because their amplitude posteriors are consistent with zero. This appears to be consistent with Fig.~9 in \cite{Giesler_19}. 

\begin{figure*}
    \centering
    \includegraphics{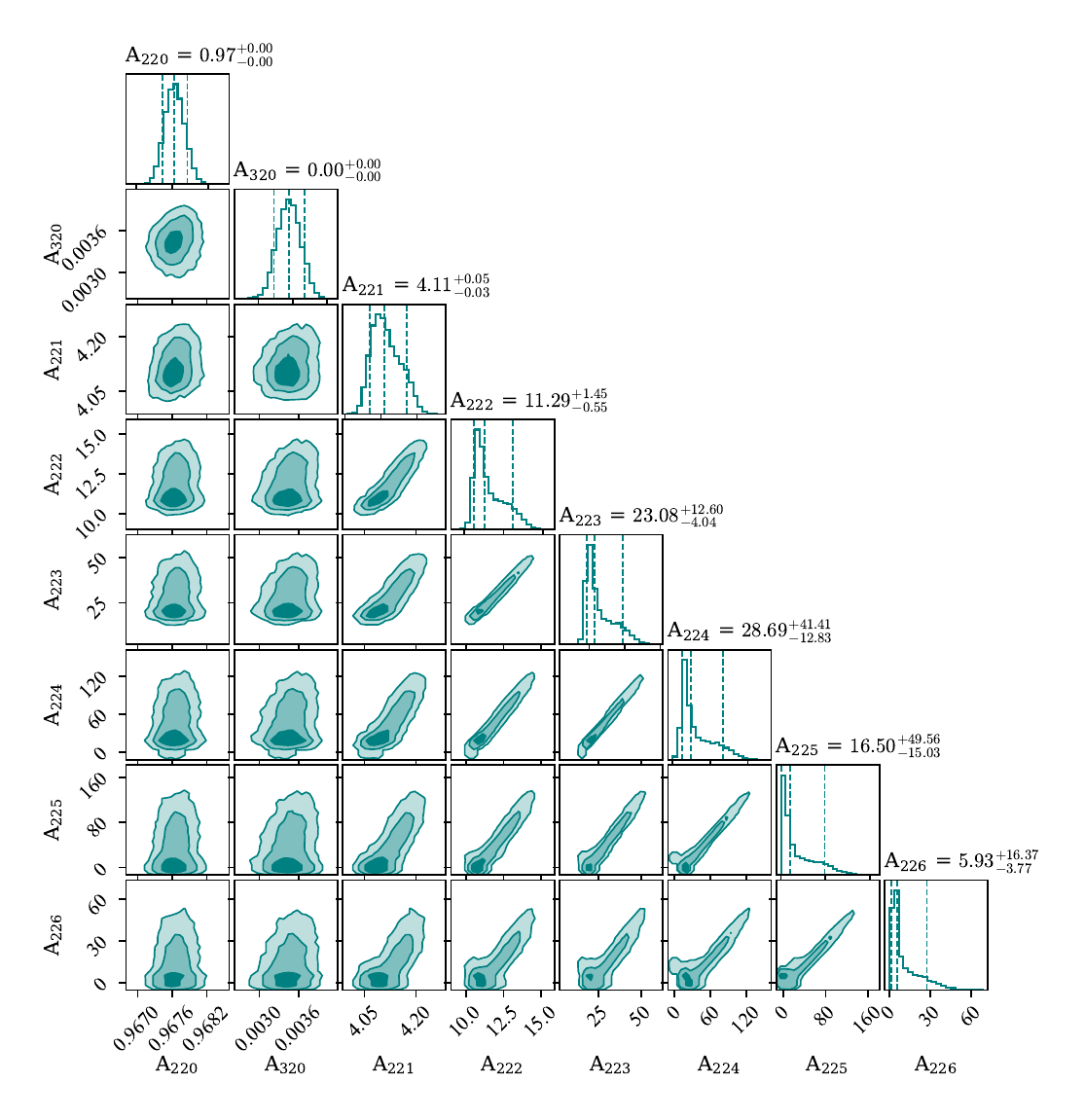}
    \caption{The amplitude posteriors for a seven-tone fit at the strain peak ($t_0 = 0M$). Tones up to $n=4$ are measured confidently with amplitudes consistent with \cite{Giesler_19}, higher overtones return posteriors that support zero. This suggests that including tones higher than the 224 may be over-fitting. We notice that the correlation between tones increases as $n$ increases. This is consistent with the findings of \cite{Bhagwat_2020}, and reflects the increasing difficulty to distinguish between higher tones that decay quickly and may not be stably recovered.}
    \label{fig:7n_corner}
\end{figure*}

We also test a model with seven tones (six tones of the 22 mode and the 320) and find that when the 225 is the highest tone, the 225 can be resolved away from zero, but is not resolved with the addition of the 226. 
It becomes consistent with zero at the next time step. 
This suggests that at least some over-fitting is likely occurring for $n=7$ within the framework of our phenomenological noise model, since those higher overtones are not resolved for this fit. 
Figure \ref{fig:7n_corner} shows the posteriors for the amplitudes of each tone included in this fit.

\section{Discussion and Conclusions}\label{sec:discussion}
In this work we introduce a framework to determine under what circumstances a perturbative description can be self-consistently applied to a binary black hole ringdown obtained from numerical relativity simulations. 
We start near the end of the waveform where the 220 tone is expected to be dominant and work backward, using the late-time waveform to set the scale of systematic uncertainty in our model.
Similarly to \cite[e.g.,][]{London_2014, Cheung_2023} we employ a criterion for stability to ensure that the fits obtained at different start times are consistent.
We find that it is possible to achieve a self-consistent perturbative description within our framework from $\approx 10 M$ after peak strain.
The perturbative description does not appear to be stable back to the peak strain given the specific set of tones considered here.

The 220 is not stably recovered without the addition of the 221 until late times. 
However, the 221 is only stably recovered and resolvable with this method for a short time interval between $10 - 35 M$.
Including the 222 appears to improve the fits at early times and allows the 220 and 221 to be recovered stably at earlier times. 
The 222 recovery can be briefly stabilised with the addition of the 223. 
However the 223 is not stably recovered at any of the times we investigate and does not seem to stabilise with the addition of the 224. 

The 320 tone, which we add as the dominant contribution from mode mixing, is always stable from the waveform peak. 
Ref.~\cite{Giesler_19} suggests that the perturbative model can be applied at or before $t_0 = t_\text{peak}$.
We show that, in our framework, at $t_0 = t_\text{peak} = 0 M$ the numerical-relativity data can be explained in a self-consistent way with a set of five tones (four tones of the 22 mode along with the 320 tone).
Any higher tones included in the fit are consistent with zero amplitude. 
None of the sets of tones beyond $N_\text{tones}=3$ include a region where \textit{all} tones included in the set are recovered stably. 
When $N_\text{tones}>3$, the highest tone transitions from unstable to unresolved with no stable region. 
If one requires that all the tones in a set are stably recovered over some finite interval, then the largest available set of tones is $N_\text{tones}=3$.

We have shown that a superposition of quasinormal tones can be used to produce a self-consistent description of the ringdown.
However, this does not prove that these fits are ``physical'' (as opposed to phenomenological) or that the signal is consistent with a perturbed black hole. 
This study brings into focus a great challenge at the heart of the black-hole spectroscopy program: it is unclear how one can even answer the question of whether the perturbative description is physical or phenomenological. The stability of the quasinormal tone fits is a requirement for their physical interpretability, but further work is required to understand the relation of these observables to the underlying nature of the spacetime and our ability to probe it.

\subsection*{Addendum}
As we were preparing this manuscript, we became aware of work by \cite{Takahashi_2023} and \cite{Qiu2023}. Ref.~\cite{Takahashi_2023} also attempts to extract tones in a self-consistent framework---by iteratively subtracting away the tone with the longest damping time. 
They find that this technique improves the stability of the extracted amplitude and phases for up to five tones of \texttt{SXS:0305}. 
This is broadly consistent with our result that only a limited number of tones can be fitted in a self-consistent framework, although we find this is limited to four tones rather than five. 
They also include the 320 contribution from mode-mixing and find that this improves the stability of the early tones but introduces more instability of tones higher than $n=2$. 
Ref.~\cite{Qiu2023} uses Bayesian inference to compare the performance of non-linear inspiral-merger-ringdown (IMR) models with linear QNM ringdown models. They find that IMR models produce better fits with higher Bayes factors at early times. They find that overtones may be measurable in high SNR events consistent with third-generation observatories. They caution that the instability of tones may make linear QNM models less reliable than non-linear models for performing tests of general relativity.

\begin{acknowledgments}
We thank the referee for their helpful suggestions to improve the manuscript. We thank Dana Jones and Gregorio Carullo for their helpful comments on this work. We also thank Saul Teukolsky, Vishal Baibhav, Will Farr, Harrison Siegel and Ben Farr for helpful advice and discussions. 
This work is supported through Australian Research Council (ARC) Centre of Excellence CE170100004, Discovery Projects DP220101610 and DP230103088, and LIEF Project LE210100002.  This work was supported in part by the Sherman Fairchild Foundation and NSF Grants No. PHY-2011968, PHY-2011961, PHY-2309211, PHY-2309231, OAC-2209656 at Caltech., as well as NSF Grants No. PHY-2207342 and OAC-2209655 at Cornell. T. A. C. receives support from the Australian Government Research Training Program. The authors are grateful for for computational resources provided by the LIGO Laboratory computing cluster at California Institute of Technology supported by National Science Foundation Grants PHY-0757058 and PHY-0823459, and the Ngarrgu Tindebeek / OzSTAR Australian national facility at Swinburne University of Technology. 
\end{acknowledgments}

\appendix
\section{Noise Model}\label{app:noise}

\begin{figure}
    \centering
    \includegraphics{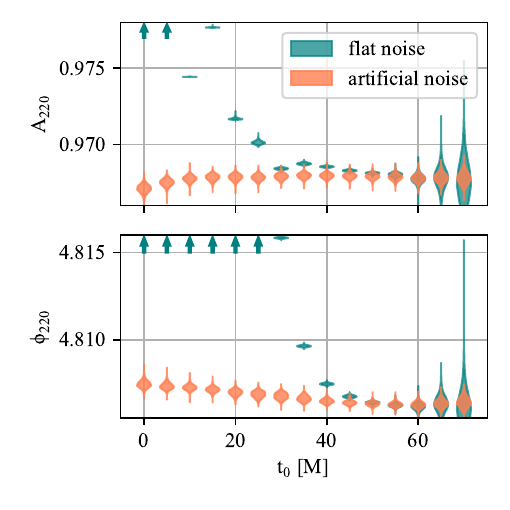}
    \caption{Posterior for amplitude and phase as a function of start time $t_0$ for the 220 tone.
    (The fits also include the 320 tone; not shown here.)
    The ``flat-noise'' model (teal) assumes the systematic error does not change with respect to $t_0$.
    The ``artificial-noise'' model  (orange) assumes that systematic error is a constant fraction of the 220 amplitude.
    The artificial-noise model produces more consistent fits than the flat-noise model.}
    \label{fig:whitening_fits}
\end{figure}

In the body of this manuscript, we employ an ``artificial noise'' model in which the systematic uncertainty is a constant error relative to the 220 tone; see Section~\ref{sec:framework}.
We also test a ``flat noise'' model in which the systematic uncertainty is constant at all times.
In Fig.~\ref{fig:whitening_fits} we show how the flat-noise fits compare to the artificial-noise fits.
(For this analysis, we include the 220 and 320 tones.)
We find that the posteriors are more comparable in magnitude and size over time using the artificial noise approach. We also find better sampler convergence, which is likely due to the SNR being comparable at early and late times rather than orders of magnitude different. 
Thus, in order to facilitate a self-consistent overtone model, we employ the artificial noise model in the main body of the manuscript. However we emphasise that this noise model is entirely phenomenological and not physically motivated. 

\section{Phase posteriors}
\label{app:phase}
\begin{figure}[ht!]
    \centering
    \includegraphics{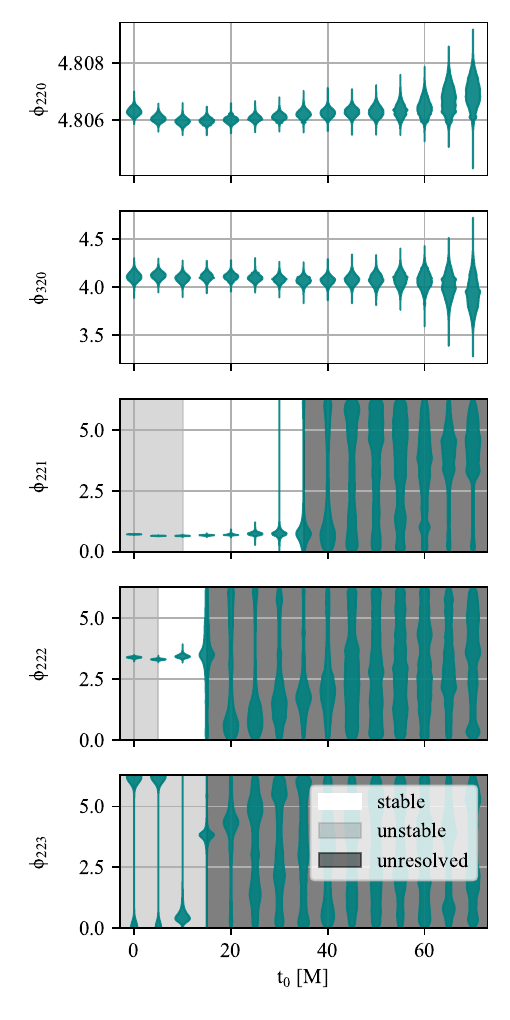}
    \caption{Posterior distributions for phase as a function of $t_0$. 
    Each panel represents a different tone.
    We show the regimes where each fit is stably recovered (white), unstably recovered (light gray) and unresolved (dark gray). 
    The 221, 222 and 223 tones are unstably recovered at early times and become unresolved at late times, which implies that they do not significantly improve the fit at late times. 
    The 223 tone is not stably recovered at any time but may stabilise with additional tones. We show the corresponding amplitude posteriors in Fig.~\ref{fig:n5_fit}. }
    \label{fig:5phi_violin}
\end{figure}

In Fig.~\ref{fig:5phi_violin} we plot the phase posteriors of the five-tone fit we present in the main body of the manuscript. The corresponding posteriors for the tone amplitudes are presented in Fig.~\ref{fig:n5_fit}. 

\section{220 stability} \label{app:stability}
\begin{figure}
   \centering
   \includegraphics{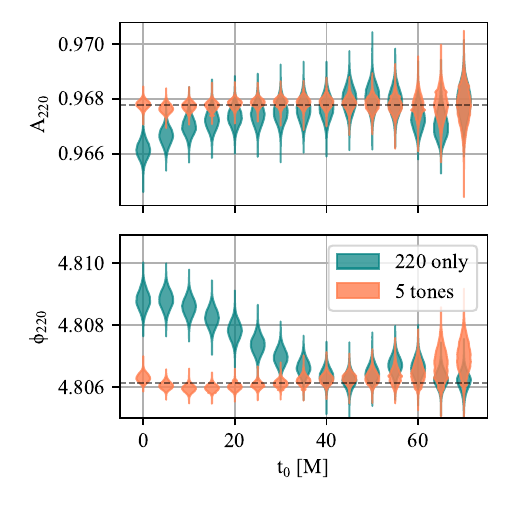}
   \caption{The amplitude and phase posteriors of the 220 tone when measured with a 1 and 5-tone model. The 5-tone model is stable at all times, while the 1-tone model becomes stable at $30M$. The dashed horizontal lines show the median of the overlap at 95\% confidence for the 5-tone model. }
   \label{fig:220_n1_n5}
\end{figure}

Figure ~\ref{fig:220_n1_n5} shows the posterior distributions recovered for the 220 tone with the lowest and highest dimensional models included in the study. We also show the median values that are common to all posteriors in the 5-tone fit. 

\section{Varying the frequency of overtones}\label{app:frequency}
\begin{figure}[h]
    \centering
    \includegraphics{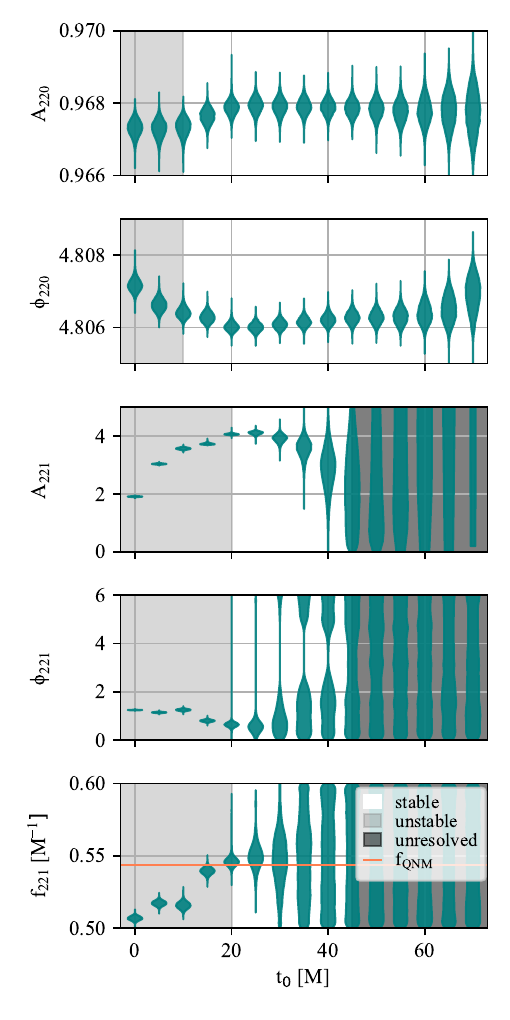}
    \caption{
    Violin plots showing the posterior distributions for a three-tone fit where $f_{221}$ is treated like a free parameter. 
    Including the 221 makes the 220 fit stable at earlier times. 
    The fit becomes stable at $20 M$, which is $10 M$ earlier than when we keep $f_{221}$ fixed. 
    Similarly, the 220 becomes stably recovered $5 M$ earlier than when $f_{221}$ was fixed. 
    }
    \label{fig:frequency}
\end{figure}

In the main body of this manuscript, we assume that the frequency and damping time of each tone are fixed to the values expected from black-hole perturbation theory.
Here, we relax this assumption so that the frequency can be treated as a free parameter.
We test a set of three tones including the 220, 320 and 221 tones and allowing the frequency of the 221 to vary. We use a uniform prior between 0.5 and 0.6 for the 221 frequency.
Figure~\ref{fig:frequency} shows the result of this fit. 
The predicted frequency value is shown as a red line in the bottom panel. 
The predicted frequency is consistent with the posterior only for start times $t_0 > 15 M$. 
The posteriors for other parameters do not change significantly when $f_{221}$ is treated as a free parameter. 
The fact that the numerical-relativity data prefer the wrong value of $f_{221}$ before $15 M$ could mean that non-linearities are present in the data at early times that are influencing the fit. 
Alternatively, it may be that more tones are required at early times to avoid this behavior, which is what \citep[e.g.,][]{Giesler_19} would imply.

We also measure the frequency of the 220 tone in a two-mode fit with the 220 and 320 tones. 
We find that the posterior moves away from the predicted frequency value while the phase becomes unstable. 
The frequency predicted by general relativity is only recovered for three time steps between 35 and $45 M$. 
In the three- and four-tone models, varying the 221 frequency slightly improves the stability of the first three tones. 
The general relativity frequency for the 221 is recovered at relatively late times: $15 M$ for a three-mode fit and $10M $ for a four-mode fit.
This further highlights that caution is required when fitting the tone frequencies due to the uncertain number of tones required, potential non-linearities and non-optimal fits, especially when doing so to test the no-hair theorem as shown in \citep[e.g.,][]{Giesler_19, isi_2021_analysing}.

Table \ref{tab:stability_with_freq} summarises the stability of the recovery of each of the tones when we sample the frequency as well as the amplitude and phase. 
Treating $f_{221}$ as a free parameter helps stabilise the fits at earlier times than when we sample in the amplitude and phase alone. 
The 221 also remains distinct from zero for an extra time step. 
However, we see the opposite effect when we treat $f_{220}$ as a free parameter for the two-tone fit.
The 220 parameters do not become stable until $60 M$ and the 320 tone is slightly affected, stabilising at $5 M$ instead of $0 M$. 
\begin{table}
    \centering
    \caption{
    The stability of our tones when we vary either the $f_{220}$ or $f_{221}$ as free parameters. 
    In the two-tone fit, we allow $f_{220}$ to vary.
  The variable $t_\text{stable}$ indicates the time at which the amplitude and phase of each tone become consistent with subsequent fits using later parts of the waveform (and which the amplitude of the tone is inconsistent with zero).
    This time (and all other times in this table are quoted in units of $M$.
    If the tone is not stably recovered for any time, this cell is marked ``$-$''.
    The variable $t_\text{unresolved}$ indicates the time at which the amplitude becomes consistent with zero.
    If the tone is never consistent with zero, this cell is marked ``$-$''. $\tau$ denotes the damping time for each tone. }
    \SetTblrInner{rowsep=0.8pt,  colsep=5pt}
    \begin{tblr}{c c c c c c}
    \hline
    \hline 
    \rule{0pt}{3ex}    
    N$_\text{tones}$ & tone & $\tau$ & t$_\text{stable}$ & t$_\text{unresolved}$ & stable interval\\
    \hline
       \SetCell[r=2]{} 2 & 220 & 11.74 & 60 & $-$ &  \SetCell[r=2]{}$60 - 70 $ \\
           & 320  & 11.27 & 5  & $-$ \\
    \hline
      \SetCell[r=3]{} 3 & 220 & 11.74 & 10 & $-$ & \SetCell[r=3]{} $20 - 45$\\
         & 320  & 11.27 &  0  & $-$\\
         & 221  & 3.88 & 20  &  45 \\
    \hline
       \SetCell[r=4]{} 4 & 220 & 11.74 & 5 & $-$ & \SetCell[r=4]{} $-$ \\
        & 320 & 11.27 & 0 & $-$ \\
        & 221 & 3.88 & 15 & 35 \\
        & 222 & 2.30 & $-$ & 20 \\
    \hline
    \hline
    \end{tblr}
 \label{tab:stability_with_freq}
\end{table}

\bibliography{bib}

\providecommand{\noopsort}[1]{}\providecommand{\singleletter}[1]{#1}%
\begin{thebibliography}{85}%
\makeatletter
\providecommand \@ifxundefined [1]{%
 \@ifx{#1\undefined}
}%
\providecommand \@ifnum [1]{%
 \ifnum #1\expandafter \@firstoftwo
 \else \expandafter \@secondoftwo
 \fi
}%
\providecommand \@ifx [1]{%
 \ifx #1\expandafter \@firstoftwo
 \else \expandafter \@secondoftwo
 \fi
}%
\providecommand \natexlab [1]{#1}%
\providecommand \enquote  [1]{``#1''}%
\providecommand \bibnamefont  [1]{#1}%
\providecommand \bibfnamefont [1]{#1}%
\providecommand \citenamefont [1]{#1}%
\providecommand \href@noop [0]{\@secondoftwo}%
\providecommand \href [0]{\begingroup \@sanitize@url \@href}%
\providecommand \@href[1]{\@@startlink{#1}\@@href}%
\providecommand \@@href[1]{\endgroup#1\@@endlink}%
\providecommand \@sanitize@url [0]{\catcode `\\12\catcode `\$12\catcode
  `\&12\catcode `\#12\catcode `\^12\catcode `\_12\catcode `\%12\relax}%
\providecommand \@@startlink[1]{}%
\providecommand \@@endlink[0]{}%
\providecommand \url  [0]{\begingroup\@sanitize@url \@url }%
\providecommand \@url [1]{\endgroup\@href {#1}{\urlprefix }}%
\providecommand \urlprefix  [0]{URL }%
\providecommand \Eprint [0]{\href }%
\providecommand \doibase [0]{https://doi.org/}%
\providecommand \selectlanguage [0]{\@gobble}%
\providecommand \bibinfo  [0]{\@secondoftwo}%
\providecommand \bibfield  [0]{\@secondoftwo}%
\providecommand \translation [1]{[#1]}%
\providecommand \BibitemOpen [0]{}%
\providecommand \bibitemStop [0]{}%
\providecommand \bibitemNoStop [0]{.\EOS\space}%
\providecommand \EOS [0]{\spacefactor3000\relax}%
\providecommand \BibitemShut  [1]{\csname bibitem#1\endcsname}%
\let\auto@bib@innerbib\@empty
\bibitem [{\citenamefont {Vishveshwara}(1970)}]{Vishveshwara_1970}%
  \BibitemOpen
  \bibfield  {author} {\bibinfo {author} {\bibfnamefont {C.~V.}\ \bibnamefont
  {Vishveshwara}},\ }\bibfield  {title} {\bibinfo {title} {Stability of the
  schwarzschild metric},\ }\href {https://doi.org/10.1103/PhysRevD.1.2870}
  {\bibfield  {journal} {\bibinfo  {journal} {Phys. Rev. D}\ }\textbf {\bibinfo
  {volume} {1}},\ \bibinfo {pages} {2870} (\bibinfo {year} {1970})}\BibitemShut
  {NoStop}%
\bibitem [{\citenamefont {{Teukolsky}}(1973)}]{Teukolsky_1973}%
  \BibitemOpen
  \bibfield  {author} {\bibinfo {author} {\bibfnamefont {S.~A.}\ \bibnamefont
  {{Teukolsky}}},\ }\bibfield  {title} {\bibinfo {title} {{Perturbations of a
  Rotating Black Hole. I. Fundamental Equations for Gravitational,
  Electromagnetic, and Neutrino-Field Perturbations}},\ }\href
  {https://doi.org/10.1086/152444} {\bibfield  {journal} {\bibinfo  {journal}
  {\apj}\ }\textbf {\bibinfo {volume} {185}},\ \bibinfo {pages} {635} (\bibinfo
  {year} {1973})}\BibitemShut {NoStop}%
\bibitem [{\citenamefont {{Chandrasekhar}}\ and\ \citenamefont
  {{Detweiler}}(1975)}]{Chandrasekhar_1975}%
  \BibitemOpen
  \bibfield  {author} {\bibinfo {author} {\bibfnamefont {S.}~\bibnamefont
  {{Chandrasekhar}}}\ and\ \bibinfo {author} {\bibfnamefont {S.}~\bibnamefont
  {{Detweiler}}},\ }\bibfield  {title} {\bibinfo {title} {{The Quasi-Normal
  Modes of the Schwarzschild Black Hole}},\ }\href
  {https://doi.org/10.1098/rspa.1975.0112} {\bibfield  {journal} {\bibinfo
  {journal} {Proceedings of the Royal Society of London Series A}\ }\textbf
  {\bibinfo {volume} {344}},\ \bibinfo {pages} {441} (\bibinfo {year}
  {1975})}\BibitemShut {NoStop}%
\bibitem [{\citenamefont {Leaver}(1985)}]{Leaver_1985}%
  \BibitemOpen
  \bibfield  {author} {\bibinfo {author} {\bibfnamefont {E.~W.}\ \bibnamefont
  {Leaver}},\ }\bibfield  {title} {\bibinfo {title} {{An Analytic
  representation for the quasi normal modes of Kerr black holes}},\ }\href
  {https://doi.org/10.1098/rspa.1985.0119} {\bibfield  {journal} {\bibinfo
  {journal} {Proc. Roy. Soc. Lond. A}\ }\textbf {\bibinfo {volume} {402}},\
  \bibinfo {pages} {285} (\bibinfo {year} {1985})}\BibitemShut {NoStop}%
\bibitem [{\citenamefont {Kokkotas}\ and\ \citenamefont
  {Schmidt}(1999)}]{Kokkotas_1999}%
  \BibitemOpen
  \bibfield  {author} {\bibinfo {author} {\bibfnamefont {K.~D.}\ \bibnamefont
  {Kokkotas}}\ and\ \bibinfo {author} {\bibfnamefont {B.~G.}\ \bibnamefont
  {Schmidt}},\ }\bibfield  {title} {\bibinfo {title} {Quasi-normal modes of
  stars and black holes},\ }\bibfield  {journal} {\bibinfo  {journal} {Living
  Reviews in Relativity}\ }\textbf {\bibinfo {volume} {2}},\ \href
  {https://doi.org/10.12942/lrr-1999-2} {10.12942/lrr-1999-2} (\bibinfo {year}
  {1999})\BibitemShut {NoStop}%
\bibitem [{\citenamefont {{Carter}}(1971)}]{Carter_1971}%
  \BibitemOpen
  \bibfield  {author} {\bibinfo {author} {\bibfnamefont {B.}~\bibnamefont
  {{Carter}}},\ }\bibfield  {title} {\bibinfo {title} {{Axisymmetric Black Hole
  Has Only Two Degrees of Freedom}},\ }\href
  {https://doi.org/10.1103/PhysRevLett.26.331} {\bibfield  {journal} {\bibinfo
  {journal} {\prl}\ }\textbf {\bibinfo {volume} {26}},\ \bibinfo {pages} {331}
  (\bibinfo {year} {1971})}\BibitemShut {NoStop}%
\bibitem [{\citenamefont {{Dreyer}}\ \emph {et~al.}(2004)\citenamefont
  {{Dreyer}}, \citenamefont {{Kelly}}, \citenamefont {{Krishnan}},
  \citenamefont {{Finn}}, \citenamefont {{Garrison}},\ and\ \citenamefont
  {{Lopez-Aleman}}}]{Dreyer2004}%
  \BibitemOpen
  \bibfield  {author} {\bibinfo {author} {\bibfnamefont {O.}~\bibnamefont
  {{Dreyer}}}, \bibinfo {author} {\bibfnamefont {B.}~\bibnamefont {{Kelly}}},
  \bibinfo {author} {\bibfnamefont {B.}~\bibnamefont {{Krishnan}}}, \bibinfo
  {author} {\bibfnamefont {L.~S.}\ \bibnamefont {{Finn}}}, \bibinfo {author}
  {\bibfnamefont {D.}~\bibnamefont {{Garrison}}},\ and\ \bibinfo {author}
  {\bibfnamefont {R.}~\bibnamefont {{Lopez-Aleman}}},\ }\bibfield  {title}
  {\bibinfo {title} {{Black-hole spectroscopy: testing general relativity
  through gravitational-wave observations}},\ }\href
  {https://doi.org/10.1088/0264-9381/21/4/003} {\bibfield  {journal} {\bibinfo
  {journal} {Classical and Quantum Gravity}\ }\textbf {\bibinfo {volume}
  {21}},\ \bibinfo {pages} {787} (\bibinfo {year} {2004})},\ \Eprint
  {https://arxiv.org/abs/gr-qc/0309007} {arXiv:gr-qc/0309007 [gr-qc]}
  \BibitemShut {NoStop}%
\bibitem [{\citenamefont {Berti}\ \emph {et~al.}(2006)\citenamefont {Berti},
  \citenamefont {Cardoso},\ and\ \citenamefont {Will}}]{Berti_2006}%
  \BibitemOpen
  \bibfield  {author} {\bibinfo {author} {\bibfnamefont {E.}~\bibnamefont
  {Berti}}, \bibinfo {author} {\bibfnamefont {V.}~\bibnamefont {Cardoso}},\
  and\ \bibinfo {author} {\bibfnamefont {C.~M.}\ \bibnamefont {Will}},\
  }\bibfield  {title} {\bibinfo {title} {Gravitational-wave spectroscopy of
  massive black holes with the space interferometer {LISA}},\ }\bibfield
  {journal} {\bibinfo  {journal} {Physical Review D}\ }\textbf {\bibinfo
  {volume} {73}},\ \href {https://doi.org/10.1103/physrevd.73.064030}
  {10.1103/physrevd.73.064030} (\bibinfo {year} {2006})\BibitemShut {NoStop}%
\bibitem [{\citenamefont {Isi}\ \emph {et~al.}(2019)\citenamefont {Isi},
  \citenamefont {Giesler}, \citenamefont {Farr}, \citenamefont {Scheel},\ and\
  \citenamefont {Teukolsky}}]{Isi_2019}%
  \BibitemOpen
  \bibfield  {author} {\bibinfo {author} {\bibfnamefont {M.}~\bibnamefont
  {Isi}}, \bibinfo {author} {\bibfnamefont {M.}~\bibnamefont {Giesler}},
  \bibinfo {author} {\bibfnamefont {W.~M.}\ \bibnamefont {Farr}}, \bibinfo
  {author} {\bibfnamefont {M.~A.}\ \bibnamefont {Scheel}},\ and\ \bibinfo
  {author} {\bibfnamefont {S.~A.}\ \bibnamefont {Teukolsky}},\ }\bibfield
  {title} {\bibinfo {title} {Testing the no-hair theorem with {GW}150914},\
  }\bibfield  {journal} {\bibinfo  {journal} {Physical Review Letters}\
  }\textbf {\bibinfo {volume} {123}},\ \href
  {https://doi.org/10.1103/physrevlett.123.111102}
  {10.1103/physrevlett.123.111102} (\bibinfo {year} {2019})\BibitemShut
  {NoStop}%
\bibitem [{\citenamefont {Abbott}\ \emph
  {et~al.}(2021{\natexlab{a}})\citenamefont {Abbott} \emph
  {et~al.}}]{LIGOScientific:2020tif}%
  \BibitemOpen
  \bibfield  {author} {\bibinfo {author} {\bibfnamefont {R.}~\bibnamefont
  {Abbott}} \emph {et~al.} (\bibinfo {collaboration} {LIGO Scientific,
  Virgo}),\ }\bibfield  {title} {\bibinfo {title} {{Tests of general relativity
  with binary black holes from the second LIGO-Virgo gravitational-wave
  transient catalog}},\ }\href {https://doi.org/10.1103/PhysRevD.103.122002}
  {\bibfield  {journal} {\bibinfo  {journal} {Phys. Rev. D}\ }\textbf {\bibinfo
  {volume} {103}},\ \bibinfo {pages} {122002} (\bibinfo {year}
  {2021}{\natexlab{a}})},\ \Eprint {https://arxiv.org/abs/2010.14529}
  {arXiv:2010.14529 [gr-qc]} \BibitemShut {NoStop}%
\bibitem [{\citenamefont {Cotesta}\ \emph {et~al.}(2022)\citenamefont
  {Cotesta}, \citenamefont {Carullo}, \citenamefont {Berti},\ and\
  \citenamefont {Cardoso}}]{Cotesta_2022}%
  \BibitemOpen
  \bibfield  {author} {\bibinfo {author} {\bibfnamefont {R.}~\bibnamefont
  {Cotesta}}, \bibinfo {author} {\bibfnamefont {G.}~\bibnamefont {Carullo}},
  \bibinfo {author} {\bibfnamefont {E.}~\bibnamefont {Berti}},\ and\ \bibinfo
  {author} {\bibfnamefont {V.}~\bibnamefont {Cardoso}},\ }\bibfield  {title}
  {\bibinfo {title} {Analysis of ringdown overtones in {GW}150914},\ }\bibfield
   {journal} {\bibinfo  {journal} {Physical Review Letters}\ }\textbf {\bibinfo
  {volume} {129}},\ \href {https://doi.org/10.1103/physrevlett.129.111102}
  {10.1103/physrevlett.129.111102} (\bibinfo {year} {2022})\BibitemShut
  {NoStop}%
\bibitem [{\citenamefont {Abbott}\ \emph
  {et~al.}(2021{\natexlab{b}})\citenamefont {Abbott} \emph
  {et~al.}}]{LIGOScientific:2021sio}%
  \BibitemOpen
  \bibfield  {author} {\bibinfo {author} {\bibfnamefont {R.}~\bibnamefont
  {Abbott}} \emph {et~al.} (\bibinfo {collaboration} {LIGO Scientific, VIRGO,
  KAGRA}),\ }\bibfield  {title} {\bibinfo {title} {{Tests of General Relativity
  with GWTC-3}},\ }\href@noop {} {\bibfield  {journal} {\bibinfo  {journal}
  {arXiv e-prints}\ } (\bibinfo {year} {2021}{\natexlab{b}})},\ \Eprint
  {https://arxiv.org/abs/2112.06861} {arXiv:2112.06861 [gr-qc]} \BibitemShut
  {NoStop}%
\bibitem [{\citenamefont {{Nollert}}(1996)}]{Nollert_1996}%
  \BibitemOpen
  \bibfield  {author} {\bibinfo {author} {\bibfnamefont {H.-P.}\ \bibnamefont
  {{Nollert}}},\ }\bibfield  {title} {\bibinfo {title} {{About the significance
  of quasinormal modes of black holes}},\ }\href
  {https://doi.org/10.1103/PhysRevD.53.4397} {\bibfield  {journal} {\bibinfo
  {journal} {\prd}\ }\textbf {\bibinfo {volume} {53}},\ \bibinfo {pages} {4397}
  (\bibinfo {year} {1996})},\ \Eprint {https://arxiv.org/abs/gr-qc/9602032}
  {arXiv:gr-qc/9602032 [gr-qc]} \BibitemShut {NoStop}%
\bibitem [{\citenamefont {{Thrane}}\ \emph {et~al.}(2017)\citenamefont
  {{Thrane}}, \citenamefont {{Lasky}},\ and\ \citenamefont
  {{Levin}}}]{Thrane_2017}%
  \BibitemOpen
  \bibfield  {author} {\bibinfo {author} {\bibfnamefont {E.}~\bibnamefont
  {{Thrane}}}, \bibinfo {author} {\bibfnamefont {P.~D.}\ \bibnamefont
  {{Lasky}}},\ and\ \bibinfo {author} {\bibfnamefont {Y.}~\bibnamefont
  {{Levin}}},\ }\bibfield  {title} {\bibinfo {title} {{Challenges for testing
  the no-hair theorem with current and planned gravitational-wave detectors}},\
  }\href {https://doi.org/10.1103/PhysRevD.96.102004} {\bibfield  {journal}
  {\bibinfo  {journal} {\prd}\ }\textbf {\bibinfo {volume} {96}},\ \bibinfo
  {eid} {102004} (\bibinfo {year} {2017})},\ \Eprint
  {https://arxiv.org/abs/1706.05152} {arXiv:1706.05152 [gr-qc]} \BibitemShut
  {NoStop}%
\bibitem [{\citenamefont {{Bhagwat}}\ \emph {et~al.}(2018)\citenamefont
  {{Bhagwat}}, \citenamefont {{Okounkova}}, \citenamefont {{Ballmer}},
  \citenamefont {{Brown}}, \citenamefont {{Giesler}}, \citenamefont
  {{Scheel}},\ and\ \citenamefont {{Teukolsky}}}]{Bhagwat_2018}%
  \BibitemOpen
  \bibfield  {author} {\bibinfo {author} {\bibfnamefont {S.}~\bibnamefont
  {{Bhagwat}}}, \bibinfo {author} {\bibfnamefont {M.}~\bibnamefont
  {{Okounkova}}}, \bibinfo {author} {\bibfnamefont {S.~W.}\ \bibnamefont
  {{Ballmer}}}, \bibinfo {author} {\bibfnamefont {D.~A.}\ \bibnamefont
  {{Brown}}}, \bibinfo {author} {\bibfnamefont {M.}~\bibnamefont {{Giesler}}},
  \bibinfo {author} {\bibfnamefont {M.~A.}\ \bibnamefont {{Scheel}}},\ and\
  \bibinfo {author} {\bibfnamefont {S.~A.}\ \bibnamefont {{Teukolsky}}},\
  }\bibfield  {title} {\bibinfo {title} {{On choosing the start time of binary
  black hole ringdowns}},\ }\href {https://doi.org/10.1103/PhysRevD.97.104065}
  {\bibfield  {journal} {\bibinfo  {journal} {\prd}\ }\textbf {\bibinfo
  {volume} {97}},\ \bibinfo {eid} {104065} (\bibinfo {year} {2018})},\ \Eprint
  {https://arxiv.org/abs/1711.00926} {arXiv:1711.00926 [gr-qc]} \BibitemShut
  {NoStop}%
\bibitem [{\citenamefont {{Isi}}\ and\ \citenamefont
  {{Farr}}(2021)}]{isi_2021_analysing}%
  \BibitemOpen
  \bibfield  {author} {\bibinfo {author} {\bibfnamefont {M.}~\bibnamefont
  {{Isi}}}\ and\ \bibinfo {author} {\bibfnamefont {W.~M.}\ \bibnamefont
  {{Farr}}},\ }\bibfield  {title} {\bibinfo {title} {{Analyzing black-hole
  ringdowns}},\ }\href {https://doi.org/10.48550/arXiv.2107.05609} {\bibfield
  {journal} {\bibinfo  {journal} {arXiv e-prints}\ ,\ \bibinfo {eid}
  {arXiv:2107.05609}} (\bibinfo {year} {2021})},\ \Eprint
  {https://arxiv.org/abs/2107.05609} {arXiv:2107.05609 [gr-qc]} \BibitemShut
  {NoStop}%
\bibitem [{\citenamefont {{Abbott}}\ \emph
  {et~al.}(2016{\natexlab{a}})\citenamefont {{Abbott}} \emph
  {et~al.}}]{TGR_2016}%
  \BibitemOpen
  \bibfield  {author} {\bibinfo {author} {\bibfnamefont {B.~P.}\ \bibnamefont
  {{Abbott}}} \emph {et~al.},\ }\bibfield  {title} {\bibinfo {title} {{Tests of
  General Relativity with GW150914}},\ }\href
  {https://doi.org/10.1103/PhysRevLett.116.221101} {\bibfield  {journal}
  {\bibinfo  {journal} {\prl}\ }\textbf {\bibinfo {volume} {116}},\ \bibinfo
  {eid} {221101} (\bibinfo {year} {2016}{\natexlab{a}})},\ \Eprint
  {https://arxiv.org/abs/1602.03841} {arXiv:1602.03841 [gr-qc]} \BibitemShut
  {NoStop}%
\bibitem [{\citenamefont {{Bhagwat}}\ \emph {et~al.}(2020)\citenamefont
  {{Bhagwat}}, \citenamefont {{Forteza}}, \citenamefont {{Pani}},\ and\
  \citenamefont {{Ferrari}}}]{Bhagwat_2020}%
  \BibitemOpen
  \bibfield  {author} {\bibinfo {author} {\bibfnamefont {S.}~\bibnamefont
  {{Bhagwat}}}, \bibinfo {author} {\bibfnamefont {X.~J.}\ \bibnamefont
  {{Forteza}}}, \bibinfo {author} {\bibfnamefont {P.}~\bibnamefont {{Pani}}},\
  and\ \bibinfo {author} {\bibfnamefont {V.}~\bibnamefont {{Ferrari}}},\
  }\bibfield  {title} {\bibinfo {title} {{Ringdown overtones, black hole
  spectroscopy, and no-hair theorem tests}},\ }\href
  {https://doi.org/10.1103/PhysRevD.101.044033} {\bibfield  {journal} {\bibinfo
   {journal} {\prd}\ }\textbf {\bibinfo {volume} {101}},\ \bibinfo {eid}
  {044033} (\bibinfo {year} {2020})},\ \Eprint
  {https://arxiv.org/abs/1910.08708} {arXiv:1910.08708 [gr-qc]} \BibitemShut
  {NoStop}%
\bibitem [{\citenamefont {{Baibhav}}\ \emph {et~al.}(2023)\citenamefont
  {{Baibhav}}, \citenamefont {{Cheung}}, \citenamefont {{Berti}}, \citenamefont
  {{Cardoso}}, \citenamefont {{Carullo}}, \citenamefont {{Cotesta}},
  \citenamefont {{Del Pozzo}},\ and\ \citenamefont {{Duque}}}]{Baibhav_2023}%
  \BibitemOpen
  \bibfield  {author} {\bibinfo {author} {\bibfnamefont {V.}~\bibnamefont
  {{Baibhav}}}, \bibinfo {author} {\bibfnamefont {M.~H.-Y.}\ \bibnamefont
  {{Cheung}}}, \bibinfo {author} {\bibfnamefont {E.}~\bibnamefont {{Berti}}},
  \bibinfo {author} {\bibfnamefont {V.}~\bibnamefont {{Cardoso}}}, \bibinfo
  {author} {\bibfnamefont {G.}~\bibnamefont {{Carullo}}}, \bibinfo {author}
  {\bibfnamefont {R.}~\bibnamefont {{Cotesta}}}, \bibinfo {author}
  {\bibfnamefont {W.}~\bibnamefont {{Del Pozzo}}},\ and\ \bibinfo {author}
  {\bibfnamefont {F.}~\bibnamefont {{Duque}}},\ }\bibfield  {title} {\bibinfo
  {title} {{Agnostic black hole spectroscopy: Quasinormal mode content of
  numerical relativity waveforms and limits of validity of linear perturbation
  theory}},\ }\href {https://doi.org/10.1103/PhysRevD.108.104020} {\bibfield
  {journal} {\bibinfo  {journal} {\prd}\ }\textbf {\bibinfo {volume} {108}},\
  \bibinfo {eid} {104020} (\bibinfo {year} {2023})},\ \Eprint
  {https://arxiv.org/abs/2302.03050} {arXiv:2302.03050 [gr-qc]} \BibitemShut
  {NoStop}%
\bibitem [{\citenamefont {{Bustillo}}\ \emph {et~al.}(2021)\citenamefont
  {{Bustillo}}, \citenamefont {{Lasky}},\ and\ \citenamefont
  {{Thrane}}}]{Bustillo_2021}%
  \BibitemOpen
  \bibfield  {author} {\bibinfo {author} {\bibfnamefont {J.~C.}\ \bibnamefont
  {{Bustillo}}}, \bibinfo {author} {\bibfnamefont {P.~D.}\ \bibnamefont
  {{Lasky}}},\ and\ \bibinfo {author} {\bibfnamefont {E.}~\bibnamefont
  {{Thrane}}},\ }\bibfield  {title} {\bibinfo {title} {{Black-hole
  spectroscopy, the no-hair theorem, and GW150914: Kerr versus Occam}},\ }\href
  {https://doi.org/10.1103/PhysRevD.103.024041} {\bibfield  {journal} {\bibinfo
   {journal} {\prd}\ }\textbf {\bibinfo {volume} {103}},\ \bibinfo {eid}
  {024041} (\bibinfo {year} {2021})}\BibitemShut {NoStop}%
\bibitem [{\citenamefont {{Hughes}}\ and\ \citenamefont
  {{Menou}}(2005)}]{Hughes2005}%
  \BibitemOpen
  \bibfield  {author} {\bibinfo {author} {\bibfnamefont {S.~A.}\ \bibnamefont
  {{Hughes}}}\ and\ \bibinfo {author} {\bibfnamefont {K.}~\bibnamefont
  {{Menou}}},\ }\bibfield  {title} {\bibinfo {title} {{Golden Binary
  Gravitational-Wave Sources: Robust Probes of Strong-Field Gravity}},\ }\href
  {https://doi.org/10.1086/428826} {\bibfield  {journal} {\bibinfo  {journal}
  {\apj}\ }\textbf {\bibinfo {volume} {623}},\ \bibinfo {pages} {689} (\bibinfo
  {year} {2005})},\ \Eprint {https://arxiv.org/abs/astro-ph/0410148}
  {arXiv:astro-ph/0410148 [astro-ph]} \BibitemShut {NoStop}%
\bibitem [{\citenamefont {Ghosh}\ \emph {et~al.}(2017)\citenamefont {Ghosh},
  \citenamefont {Johnson-McDaniel}, \citenamefont {Ghosh}, \citenamefont
  {Mishra}, \citenamefont {Ajith}, \citenamefont {Pozzo}, \citenamefont
  {Berry}, \citenamefont {Nielsen},\ and\ \citenamefont {London}}]{Ghosh_2017}%
  \BibitemOpen
  \bibfield  {author} {\bibinfo {author} {\bibfnamefont {A.}~\bibnamefont
  {Ghosh}}, \bibinfo {author} {\bibfnamefont {N.~K.}\ \bibnamefont
  {Johnson-McDaniel}}, \bibinfo {author} {\bibfnamefont {A.}~\bibnamefont
  {Ghosh}}, \bibinfo {author} {\bibfnamefont {C.~K.}\ \bibnamefont {Mishra}},
  \bibinfo {author} {\bibfnamefont {P.}~\bibnamefont {Ajith}}, \bibinfo
  {author} {\bibfnamefont {W.~D.}\ \bibnamefont {Pozzo}}, \bibinfo {author}
  {\bibfnamefont {C.~P.~L.}\ \bibnamefont {Berry}}, \bibinfo {author}
  {\bibfnamefont {A.~B.}\ \bibnamefont {Nielsen}},\ and\ \bibinfo {author}
  {\bibfnamefont {L.}~\bibnamefont {London}},\ }\bibfield  {title} {\bibinfo
  {title} {Testing general relativity using gravitational wave signals from the
  inspiral, merger and ringdown of binary black holes},\ }\href
  {https://doi.org/10.1088/1361-6382/aa972e} {\bibfield  {journal} {\bibinfo
  {journal} {Classical and Quantum Gravity}\ }\textbf {\bibinfo {volume}
  {35}},\ \bibinfo {pages} {014002} (\bibinfo {year} {2017})}\BibitemShut
  {NoStop}%
\bibitem [{\citenamefont {{Johnson-McDaniel}}\ \emph
  {et~al.}(2022)\citenamefont {{Johnson-McDaniel}}, \citenamefont {{Ghosh}},
  \citenamefont {{Ghonge}}, \citenamefont {{Saleem}}, \citenamefont
  {{Krishnendu}},\ and\ \citenamefont {{Clark}}}]{Johnson-McDaniel2022}%
  \BibitemOpen
  \bibfield  {author} {\bibinfo {author} {\bibfnamefont {N.~K.}\ \bibnamefont
  {{Johnson-McDaniel}}}, \bibinfo {author} {\bibfnamefont {A.}~\bibnamefont
  {{Ghosh}}}, \bibinfo {author} {\bibfnamefont {S.}~\bibnamefont {{Ghonge}}},
  \bibinfo {author} {\bibfnamefont {M.}~\bibnamefont {{Saleem}}}, \bibinfo
  {author} {\bibfnamefont {N.~V.}\ \bibnamefont {{Krishnendu}}},\ and\ \bibinfo
  {author} {\bibfnamefont {J.~A.}\ \bibnamefont {{Clark}}},\ }\bibfield
  {title} {\bibinfo {title} {{Investigating the relation between gravitational
  wave tests of general relativity}},\ }\href
  {https://doi.org/10.1103/PhysRevD.105.044020} {\bibfield  {journal} {\bibinfo
   {journal} {\prd}\ }\textbf {\bibinfo {volume} {105}},\ \bibinfo {eid}
  {044020} (\bibinfo {year} {2022})},\ \Eprint
  {https://arxiv.org/abs/2109.06988} {arXiv:2109.06988 [gr-qc]} \BibitemShut
  {NoStop}%
\bibitem [{\citenamefont {{Brito}}\ \emph {et~al.}(2018)\citenamefont
  {{Brito}}, \citenamefont {{Buonanno}},\ and\ \citenamefont
  {{Raymond}}}]{Brito2018}%
  \BibitemOpen
  \bibfield  {author} {\bibinfo {author} {\bibfnamefont {R.}~\bibnamefont
  {{Brito}}}, \bibinfo {author} {\bibfnamefont {A.}~\bibnamefont
  {{Buonanno}}},\ and\ \bibinfo {author} {\bibfnamefont {V.}~\bibnamefont
  {{Raymond}}},\ }\bibfield  {title} {\bibinfo {title} {{Black-hole
  spectroscopy by making full use of gravitational-wave modeling}},\ }\href
  {https://doi.org/10.1103/PhysRevD.98.084038} {\bibfield  {journal} {\bibinfo
  {journal} {\prd}\ }\textbf {\bibinfo {volume} {98}},\ \bibinfo {eid} {084038}
  (\bibinfo {year} {2018})},\ \Eprint {https://arxiv.org/abs/1805.00293}
  {arXiv:1805.00293 [gr-qc]} \BibitemShut {NoStop}%
\bibitem [{\citenamefont {{Gennari}}\ \emph {et~al.}(2023)\citenamefont
  {{Gennari}}, \citenamefont {{Carullo}},\ and\ \citenamefont {{Del
  Pozzo}}}]{Gennari2023}%
  \BibitemOpen
  \bibfield  {author} {\bibinfo {author} {\bibfnamefont {V.}~\bibnamefont
  {{Gennari}}}, \bibinfo {author} {\bibfnamefont {G.}~\bibnamefont
  {{Carullo}}},\ and\ \bibinfo {author} {\bibfnamefont {W.}~\bibnamefont {{Del
  Pozzo}}},\ }\bibfield  {title} {\bibinfo {title} {{Searching for ringdown
  higher modes with a numerical relativity-informed post-merger model}},\
  }\href {https://doi.org/10.48550/arXiv.2312.12515} {\bibfield  {journal}
  {\bibinfo  {journal} {arXiv e-prints}\ ,\ \bibinfo {eid} {arXiv:2312.12515}}
  (\bibinfo {year} {2023})},\ \Eprint {https://arxiv.org/abs/2312.12515}
  {arXiv:2312.12515 [gr-qc]} \BibitemShut {NoStop}%
\bibitem [{\citenamefont {{Pretorius}}(2005)}]{Pretorius_2005}%
  \BibitemOpen
  \bibfield  {author} {\bibinfo {author} {\bibfnamefont {F.}~\bibnamefont
  {{Pretorius}}},\ }\bibfield  {title} {\bibinfo {title} {{Numerical relativity
  using a generalized harmonic decomposition}},\ }\href
  {https://doi.org/10.1088/0264-9381/22/2/014} {\bibfield  {journal} {\bibinfo
  {journal} {Classical and Quantum Gravity}\ }\textbf {\bibinfo {volume}
  {22}},\ \bibinfo {pages} {425} (\bibinfo {year} {2005})},\ \Eprint
  {https://arxiv.org/abs/gr-qc/0407110} {arXiv:gr-qc/0407110 [gr-qc]}
  \BibitemShut {NoStop}%
\bibitem [{\citenamefont {{Campanelli}}\ \emph {et~al.}(2006)\citenamefont
  {{Campanelli}}, \citenamefont {{Lousto}}, \citenamefont {{Marronetti}},\ and\
  \citenamefont {{Zlochower}}}]{Campanelli_2006}%
  \BibitemOpen
  \bibfield  {author} {\bibinfo {author} {\bibfnamefont {M.}~\bibnamefont
  {{Campanelli}}}, \bibinfo {author} {\bibfnamefont {C.~O.}\ \bibnamefont
  {{Lousto}}}, \bibinfo {author} {\bibfnamefont {P.}~\bibnamefont
  {{Marronetti}}},\ and\ \bibinfo {author} {\bibfnamefont {Y.}~\bibnamefont
  {{Zlochower}}},\ }\bibfield  {title} {\bibinfo {title} {{Accurate Evolutions
  of Orbiting Black-Hole Binaries without Excision}},\ }\href
  {https://doi.org/10.1103/PhysRevLett.96.111101} {\bibfield  {journal}
  {\bibinfo  {journal} {\prl}\ }\textbf {\bibinfo {volume} {96}},\ \bibinfo
  {eid} {111101} (\bibinfo {year} {2006})},\ \Eprint
  {https://arxiv.org/abs/gr-qc/0511048} {arXiv:gr-qc/0511048 [gr-qc]}
  \BibitemShut {NoStop}%
\bibitem [{\citenamefont {Baker}\ \emph {et~al.}(2006)\citenamefont {Baker},
  \citenamefont {Centrella}, \citenamefont {Choi}, \citenamefont {Koppitz},\
  and\ \citenamefont {van Meter}}]{Baker_2006}%
  \BibitemOpen
  \bibfield  {author} {\bibinfo {author} {\bibfnamefont {J.~G.}\ \bibnamefont
  {Baker}}, \bibinfo {author} {\bibfnamefont {J.}~\bibnamefont {Centrella}},
  \bibinfo {author} {\bibfnamefont {D.-I.}\ \bibnamefont {Choi}}, \bibinfo
  {author} {\bibfnamefont {M.}~\bibnamefont {Koppitz}},\ and\ \bibinfo {author}
  {\bibfnamefont {J.}~\bibnamefont {van Meter}},\ }\bibfield  {title} {\bibinfo
  {title} {Gravitational-wave extraction from an inspiraling configuration of
  merging black holes},\ }\bibfield  {journal} {\bibinfo  {journal} {Physical
  Review Letters}\ }\textbf {\bibinfo {volume} {96}},\ \href
  {https://doi.org/10.1103/physrevlett.96.111102}
  {10.1103/physrevlett.96.111102} (\bibinfo {year} {2006})\BibitemShut
  {NoStop}%
\bibitem [{\citenamefont {{Buonanno}}\ \emph {et~al.}(2007)\citenamefont
  {{Buonanno}}, \citenamefont {{Cook}},\ and\ \citenamefont
  {{Pretorius}}}]{Buonanno_2007}%
  \BibitemOpen
  \bibfield  {author} {\bibinfo {author} {\bibfnamefont {A.}~\bibnamefont
  {{Buonanno}}}, \bibinfo {author} {\bibfnamefont {G.~B.}\ \bibnamefont
  {{Cook}}},\ and\ \bibinfo {author} {\bibfnamefont {F.}~\bibnamefont
  {{Pretorius}}},\ }\bibfield  {title} {\bibinfo {title} {{Inspiral, merger,
  and ring-down of equal-mass black-hole binaries}},\ }\href
  {https://doi.org/10.1103/PhysRevD.75.124018} {\bibfield  {journal} {\bibinfo
  {journal} {\prd}\ }\textbf {\bibinfo {volume} {75}},\ \bibinfo {eid} {124018}
  (\bibinfo {year} {2007})},\ \Eprint {https://arxiv.org/abs/gr-qc/0610122}
  {arXiv:gr-qc/0610122 [gr-qc]} \BibitemShut {NoStop}%
\bibitem [{\citenamefont {{Baibhav}}\ \emph {et~al.}(2018)\citenamefont
  {{Baibhav}}, \citenamefont {{Berti}}, \citenamefont {{Cardoso}},\ and\
  \citenamefont {{Khanna}}}]{Baibhav_2018}%
  \BibitemOpen
  \bibfield  {author} {\bibinfo {author} {\bibfnamefont {V.}~\bibnamefont
  {{Baibhav}}}, \bibinfo {author} {\bibfnamefont {E.}~\bibnamefont {{Berti}}},
  \bibinfo {author} {\bibfnamefont {V.}~\bibnamefont {{Cardoso}}},\ and\
  \bibinfo {author} {\bibfnamefont {G.}~\bibnamefont {{Khanna}}},\ }\bibfield
  {title} {\bibinfo {title} {{Black hole spectroscopy: Systematic errors and
  ringdown energy estimates}},\ }\href
  {https://doi.org/10.1103/PhysRevD.97.044048} {\bibfield  {journal} {\bibinfo
  {journal} {\prd}\ }\textbf {\bibinfo {volume} {97}},\ \bibinfo {eid} {044048}
  (\bibinfo {year} {2018})},\ \Eprint {https://arxiv.org/abs/1710.02156}
  {arXiv:1710.02156 [gr-qc]} \BibitemShut {NoStop}%
\bibitem [{\citenamefont {{Ota}}\ and\ \citenamefont
  {{Chirenti}}(2020)}]{Ota2020}%
  \BibitemOpen
  \bibfield  {author} {\bibinfo {author} {\bibfnamefont {I.}~\bibnamefont
  {{Ota}}}\ and\ \bibinfo {author} {\bibfnamefont {C.}~\bibnamefont
  {{Chirenti}}},\ }\bibfield  {title} {\bibinfo {title} {{Overtones or higher
  harmonics? Prospects for testing the no-hair theorem with gravitational wave
  detections}},\ }\href {https://doi.org/10.1103/PhysRevD.101.104005}
  {\bibfield  {journal} {\bibinfo  {journal} {\prd}\ }\textbf {\bibinfo
  {volume} {101}},\ \bibinfo {eid} {104005} (\bibinfo {year} {2020})},\ \Eprint
  {https://arxiv.org/abs/1911.00440} {arXiv:1911.00440 [gr-qc]} \BibitemShut
  {NoStop}%
\bibitem [{\citenamefont {{Mourier}}\ \emph {et~al.}(2021)\citenamefont
  {{Mourier}}, \citenamefont {{Jim{\'e}nez Forteza}}, \citenamefont
  {{Pook-Kolb}}, \citenamefont {{Krishnan}},\ and\ \citenamefont
  {{Schnetter}}}]{Mourier2021}%
  \BibitemOpen
  \bibfield  {author} {\bibinfo {author} {\bibfnamefont {P.}~\bibnamefont
  {{Mourier}}}, \bibinfo {author} {\bibfnamefont {X.}~\bibnamefont
  {{Jim{\'e}nez Forteza}}}, \bibinfo {author} {\bibfnamefont {D.}~\bibnamefont
  {{Pook-Kolb}}}, \bibinfo {author} {\bibfnamefont {B.}~\bibnamefont
  {{Krishnan}}},\ and\ \bibinfo {author} {\bibfnamefont {E.}~\bibnamefont
  {{Schnetter}}},\ }\bibfield  {title} {\bibinfo {title} {{Quasinormal modes
  and their overtones at the common horizon in a binary black hole merger}},\
  }\href {https://doi.org/10.1103/PhysRevD.103.044054} {\bibfield  {journal}
  {\bibinfo  {journal} {\prd}\ }\textbf {\bibinfo {volume} {103}},\ \bibinfo
  {eid} {044054} (\bibinfo {year} {2021})},\ \Eprint
  {https://arxiv.org/abs/2010.15186} {arXiv:2010.15186 [gr-qc]} \BibitemShut
  {NoStop}%
\bibitem [{\citenamefont {{Sago}}\ \emph {et~al.}(2021)\citenamefont {{Sago}},
  \citenamefont {{Isoyama}},\ and\ \citenamefont {{Nakano}}}]{Sago2021}%
  \BibitemOpen
  \bibfield  {author} {\bibinfo {author} {\bibfnamefont {N.}~\bibnamefont
  {{Sago}}}, \bibinfo {author} {\bibfnamefont {S.}~\bibnamefont {{Isoyama}}},\
  and\ \bibinfo {author} {\bibfnamefont {H.}~\bibnamefont {{Nakano}}},\
  }\bibfield  {title} {\bibinfo {title} {{Fundamental Tone and Overtones of
  Quasinormal Modes in Ringdown Gravitational Waves: A Detailed Study in Black
  Hole Perturbation}},\ }\href {https://doi.org/10.3390/universe7100357}
  {\bibfield  {journal} {\bibinfo  {journal} {Universe}\ }\textbf {\bibinfo
  {volume} {7}},\ \bibinfo {eid} {357} (\bibinfo {year} {2021})},\ \Eprint
  {https://arxiv.org/abs/2108.13017} {arXiv:2108.13017 [gr-qc]} \BibitemShut
  {NoStop}%
\bibitem [{\citenamefont {{Li}}\ \emph {et~al.}(2022)\citenamefont {{Li}},
  \citenamefont {{Sun}}, \citenamefont {{Lo}}, \citenamefont {{Payne}},\ and\
  \citenamefont {{Chen}}}]{Li_2022}%
  \BibitemOpen
  \bibfield  {author} {\bibinfo {author} {\bibfnamefont {X.}~\bibnamefont
  {{Li}}}, \bibinfo {author} {\bibfnamefont {L.}~\bibnamefont {{Sun}}},
  \bibinfo {author} {\bibfnamefont {R.~K.~L.}\ \bibnamefont {{Lo}}}, \bibinfo
  {author} {\bibfnamefont {E.}~\bibnamefont {{Payne}}},\ and\ \bibinfo {author}
  {\bibfnamefont {Y.}~\bibnamefont {{Chen}}},\ }\bibfield  {title} {\bibinfo
  {title} {{Angular emission patterns of remnant black holes}},\ }\href
  {https://doi.org/10.1103/PhysRevD.105.024016} {\bibfield  {journal} {\bibinfo
   {journal} {\prd}\ }\textbf {\bibinfo {volume} {105}},\ \bibinfo {eid}
  {024016} (\bibinfo {year} {2022})},\ \Eprint
  {https://arxiv.org/abs/2110.03116} {arXiv:2110.03116 [gr-qc]} \BibitemShut
  {NoStop}%
\bibitem [{\citenamefont {{Giesler}}\ \emph {et~al.}(2019)\citenamefont
  {{Giesler}}, \citenamefont {{Isi}}, \citenamefont {{Scheel}},\ and\
  \citenamefont {{Teukolsky}}}]{Giesler_19}%
  \BibitemOpen
  \bibfield  {author} {\bibinfo {author} {\bibfnamefont {M.}~\bibnamefont
  {{Giesler}}}, \bibinfo {author} {\bibfnamefont {M.}~\bibnamefont {{Isi}}},
  \bibinfo {author} {\bibfnamefont {M.~A.}\ \bibnamefont {{Scheel}}},\ and\
  \bibinfo {author} {\bibfnamefont {S.~A.}\ \bibnamefont {{Teukolsky}}},\
  }\bibfield  {title} {\bibinfo {title} {{Black Hole Ringdown: The Importance
  of Overtones}},\ }\href {https://doi.org/10.1103/PhysRevX.9.041060}
  {\bibfield  {journal} {\bibinfo  {journal} {Physical Review X}\ }\textbf
  {\bibinfo {volume} {9}},\ \bibinfo {eid} {041060} (\bibinfo {year} {2019})},\
  \Eprint {https://arxiv.org/abs/1903.08284} {arXiv:1903.08284 [gr-qc]}
  \BibitemShut {NoStop}%
\bibitem [{\citenamefont {{Ma}}\ \emph {et~al.}(2022)\citenamefont {{Ma}},
  \citenamefont {{Mitman}}, \citenamefont {{Sun}}, \citenamefont {{Deppe}},
  \citenamefont {{H{\'e}bert}}, \citenamefont {{Kidder}}, \citenamefont
  {{Moxon}}, \citenamefont {{Throwe}}, \citenamefont {{Vu}},\ and\
  \citenamefont {{Chen}}}]{Ma_2022}%
  \BibitemOpen
  \bibfield  {author} {\bibinfo {author} {\bibfnamefont {S.}~\bibnamefont
  {{Ma}}}, \bibinfo {author} {\bibfnamefont {K.}~\bibnamefont {{Mitman}}},
  \bibinfo {author} {\bibfnamefont {L.}~\bibnamefont {{Sun}}}, \bibinfo
  {author} {\bibfnamefont {N.}~\bibnamefont {{Deppe}}}, \bibinfo {author}
  {\bibfnamefont {F.}~\bibnamefont {{H{\'e}bert}}}, \bibinfo {author}
  {\bibfnamefont {L.~E.}\ \bibnamefont {{Kidder}}}, \bibinfo {author}
  {\bibfnamefont {J.}~\bibnamefont {{Moxon}}}, \bibinfo {author} {\bibfnamefont
  {W.}~\bibnamefont {{Throwe}}}, \bibinfo {author} {\bibfnamefont {N.~L.}\
  \bibnamefont {{Vu}}},\ and\ \bibinfo {author} {\bibfnamefont
  {Y.}~\bibnamefont {{Chen}}},\ }\bibfield  {title} {\bibinfo {title}
  {{Quasinormal-mode filters: A new approach to analyze the gravitational-wave
  ringdown of binary black-hole mergers}},\ }\href
  {https://doi.org/10.1103/PhysRevD.106.084036} {\bibfield  {journal} {\bibinfo
   {journal} {\prd}\ }\textbf {\bibinfo {volume} {106}},\ \bibinfo {eid}
  {084036} (\bibinfo {year} {2022})},\ \Eprint
  {https://arxiv.org/abs/2207.10870} {arXiv:2207.10870 [gr-qc]} \BibitemShut
  {NoStop}%
\bibitem [{\citenamefont {{Cook}}(2020)}]{Cook2020}%
  \BibitemOpen
  \bibfield  {author} {\bibinfo {author} {\bibfnamefont {G.~B.}\ \bibnamefont
  {{Cook}}},\ }\bibfield  {title} {\bibinfo {title} {{Aspects of multimode Kerr
  ringdown fitting}},\ }\href {https://doi.org/10.1103/PhysRevD.102.024027}
  {\bibfield  {journal} {\bibinfo  {journal} {\prd}\ }\textbf {\bibinfo
  {volume} {102}},\ \bibinfo {eid} {024027} (\bibinfo {year} {2020})},\ \Eprint
  {https://arxiv.org/abs/2004.08347} {arXiv:2004.08347 [gr-qc]} \BibitemShut
  {NoStop}%
\bibitem [{\citenamefont {{Zertuche}}\ \emph {et~al.}(2022)\citenamefont
  {{Zertuche}}, \citenamefont {{Mitman}}, \citenamefont {{Khera}},
  \citenamefont {{Stein}}, \citenamefont {{Boyle}}, \citenamefont {{Deppe}},
  \citenamefont {{H{\'e}bert}}, \citenamefont {{Iozzo}}, \citenamefont
  {{Kidder}}, \citenamefont {{Moxon}}, \citenamefont {{Pfeiffer}},
  \citenamefont {{Scheel}}, \citenamefont {{Teukolsky}}, \citenamefont
  {{Throwe}},\ and\ \citenamefont {{Vu}}}]{Zertuche_2022}%
  \BibitemOpen
  \bibfield  {author} {\bibinfo {author} {\bibfnamefont {L.~M.}\ \bibnamefont
  {{Zertuche}}}, \bibinfo {author} {\bibfnamefont {K.}~\bibnamefont
  {{Mitman}}}, \bibinfo {author} {\bibfnamefont {N.}~\bibnamefont {{Khera}}},
  \bibinfo {author} {\bibfnamefont {L.~C.}\ \bibnamefont {{Stein}}}, \bibinfo
  {author} {\bibfnamefont {M.}~\bibnamefont {{Boyle}}}, \bibinfo {author}
  {\bibfnamefont {N.}~\bibnamefont {{Deppe}}}, \bibinfo {author} {\bibfnamefont
  {F.}~\bibnamefont {{H{\'e}bert}}}, \bibinfo {author} {\bibfnamefont
  {D.~A.~B.}\ \bibnamefont {{Iozzo}}}, \bibinfo {author} {\bibfnamefont
  {L.~E.}\ \bibnamefont {{Kidder}}}, \bibinfo {author} {\bibfnamefont
  {J.}~\bibnamefont {{Moxon}}}, \bibinfo {author} {\bibfnamefont {H.~P.}\
  \bibnamefont {{Pfeiffer}}}, \bibinfo {author} {\bibfnamefont {M.~A.}\
  \bibnamefont {{Scheel}}}, \bibinfo {author} {\bibfnamefont {S.~A.}\
  \bibnamefont {{Teukolsky}}}, \bibinfo {author} {\bibfnamefont
  {W.}~\bibnamefont {{Throwe}}},\ and\ \bibinfo {author} {\bibfnamefont
  {N.}~\bibnamefont {{Vu}}},\ }\bibfield  {title} {\bibinfo {title} {{High
  precision ringdown modeling: Multimode fits and BMS frames}},\ }\href
  {https://doi.org/10.1103/PhysRevD.105.104015} {\bibfield  {journal} {\bibinfo
   {journal} {\prd}\ }\textbf {\bibinfo {volume} {105}},\ \bibinfo {eid}
  {104015} (\bibinfo {year} {2022})}\BibitemShut {NoStop}%
\bibitem [{\citenamefont {{Forteza}}\ and\ \citenamefont
  {{Mourier}}(2021)}]{Forteza_2021}%
  \BibitemOpen
  \bibfield  {author} {\bibinfo {author} {\bibfnamefont {X.~J.}\ \bibnamefont
  {{Forteza}}}\ and\ \bibinfo {author} {\bibfnamefont {P.}~\bibnamefont
  {{Mourier}}},\ }\bibfield  {title} {\bibinfo {title} {{High-overtone fits to
  numerical relativity ringdowns: Beyond the dismissed n =8 special tone}},\
  }\href {https://doi.org/10.1103/PhysRevD.104.124072} {\bibfield  {journal}
  {\bibinfo  {journal} {\prd}\ }\textbf {\bibinfo {volume} {104}},\ \bibinfo
  {eid} {124072} (\bibinfo {year} {2021})},\ \Eprint
  {https://arxiv.org/abs/2107.11829} {arXiv:2107.11829 [gr-qc]} \BibitemShut
  {NoStop}%
\bibitem [{\citenamefont {{Okounkova}}(2020)}]{Okounkova_2020}%
  \BibitemOpen
  \bibfield  {author} {\bibinfo {author} {\bibfnamefont {M.}~\bibnamefont
  {{Okounkova}}},\ }\bibfield  {title} {\bibinfo {title} {{Revisiting
  non-linearity in binary black hole mergers}},\ }\href
  {https://doi.org/10.48550/arXiv.2004.00671} {\bibfield  {journal} {\bibinfo
  {journal} {arXiv e-prints}\ ,\ \bibinfo {eid} {arXiv:2004.00671}} (\bibinfo
  {year} {2020})},\ \Eprint {https://arxiv.org/abs/2004.00671}
  {arXiv:2004.00671 [gr-qc]} \BibitemShut {NoStop}%
\bibitem [{\citenamefont {Chen}\ \emph {et~al.}(2022)\citenamefont {Chen},
  \citenamefont {Kumar}, \citenamefont {Khera}, \citenamefont {Deppe},
  \citenamefont {Dhani}, \citenamefont {Boyle}, \citenamefont {Giesler},
  \citenamefont {Kidder}, \citenamefont {Pfeiffer}, \citenamefont {Scheel},\
  and\ \citenamefont {Teukolsky}}]{Chen_2022}%
  \BibitemOpen
  \bibfield  {author} {\bibinfo {author} {\bibfnamefont {Y.}~\bibnamefont
  {Chen}}, \bibinfo {author} {\bibfnamefont {P.}~\bibnamefont {Kumar}},
  \bibinfo {author} {\bibfnamefont {N.}~\bibnamefont {Khera}}, \bibinfo
  {author} {\bibfnamefont {N.}~\bibnamefont {Deppe}}, \bibinfo {author}
  {\bibfnamefont {A.}~\bibnamefont {Dhani}}, \bibinfo {author} {\bibfnamefont
  {M.}~\bibnamefont {Boyle}}, \bibinfo {author} {\bibfnamefont
  {M.}~\bibnamefont {Giesler}}, \bibinfo {author} {\bibfnamefont {L.~E.}\
  \bibnamefont {Kidder}}, \bibinfo {author} {\bibfnamefont {H.~P.}\
  \bibnamefont {Pfeiffer}}, \bibinfo {author} {\bibfnamefont {M.~A.}\
  \bibnamefont {Scheel}},\ and\ \bibinfo {author} {\bibfnamefont {S.~A.}\
  \bibnamefont {Teukolsky}},\ }\bibfield  {title} {\bibinfo {title} {Multipole
  moments on the common horizon in a binary-black-hole simulation},\ }\bibfield
   {journal} {\bibinfo  {journal} {Physical Review D}\ }\textbf {\bibinfo
  {volume} {106}},\ \href {https://doi.org/10.1103/physrevd.106.124045}
  {10.1103/physrevd.106.124045} (\bibinfo {year} {2022})\BibitemShut {NoStop}%
\bibitem [{\citenamefont {{Mitman}}\ \emph {et~al.}(2023)\citenamefont
  {{Mitman}}, \citenamefont {{Lagos}}, \citenamefont {{Stein}}, \citenamefont
  {{Ma}}, \citenamefont {{Hui}}, \citenamefont {{Chen}}, \citenamefont
  {{Deppe}}, \citenamefont {{H{\'e}bert}}, \citenamefont {{Kidder}},
  \citenamefont {{Moxon}}, \citenamefont {{Scheel}}, \citenamefont
  {{Teukolsky}}, \citenamefont {{Throwe}},\ and\ \citenamefont
  {{Vu}}}]{Mitman_2023}%
  \BibitemOpen
  \bibfield  {author} {\bibinfo {author} {\bibfnamefont {K.}~\bibnamefont
  {{Mitman}}}, \bibinfo {author} {\bibfnamefont {M.}~\bibnamefont {{Lagos}}},
  \bibinfo {author} {\bibfnamefont {L.~C.}\ \bibnamefont {{Stein}}}, \bibinfo
  {author} {\bibfnamefont {S.}~\bibnamefont {{Ma}}}, \bibinfo {author}
  {\bibfnamefont {L.}~\bibnamefont {{Hui}}}, \bibinfo {author} {\bibfnamefont
  {Y.}~\bibnamefont {{Chen}}}, \bibinfo {author} {\bibfnamefont
  {N.}~\bibnamefont {{Deppe}}}, \bibinfo {author} {\bibfnamefont
  {F.}~\bibnamefont {{H{\'e}bert}}}, \bibinfo {author} {\bibfnamefont {L.~E.}\
  \bibnamefont {{Kidder}}}, \bibinfo {author} {\bibfnamefont {J.}~\bibnamefont
  {{Moxon}}}, \bibinfo {author} {\bibfnamefont {M.~A.}\ \bibnamefont
  {{Scheel}}}, \bibinfo {author} {\bibfnamefont {S.~A.}\ \bibnamefont
  {{Teukolsky}}}, \bibinfo {author} {\bibfnamefont {W.}~\bibnamefont
  {{Throwe}}},\ and\ \bibinfo {author} {\bibfnamefont {N.~L.}\ \bibnamefont
  {{Vu}}},\ }\bibfield  {title} {\bibinfo {title} {{Nonlinearities in Black
  Hole Ringdowns}},\ }\href {https://doi.org/10.1103/PhysRevLett.130.081402}
  {\bibfield  {journal} {\bibinfo  {journal} {\prl}\ }\textbf {\bibinfo
  {volume} {130}},\ \bibinfo {eid} {081402} (\bibinfo {year} {2023})},\ \Eprint
  {https://arxiv.org/abs/2208.07380} {arXiv:2208.07380 [gr-qc]} \BibitemShut
  {NoStop}%
\bibitem [{\citenamefont {Cheung}\ \emph {et~al.}(2023)\citenamefont {Cheung},
  \citenamefont {Baibhav}, \citenamefont {Berti}, \citenamefont {Cardoso},
  \citenamefont {Carullo}, \citenamefont {Cotesta}, \citenamefont {Del~Pozzo},
  \citenamefont {Duque}, \citenamefont {Helfer}, \citenamefont {Shukla},\ and\
  \citenamefont {Wong}}]{Cheung_2023a}%
  \BibitemOpen
  \bibfield  {author} {\bibinfo {author} {\bibfnamefont {M.~H.-Y.}\
  \bibnamefont {Cheung}}, \bibinfo {author} {\bibfnamefont {V.}~\bibnamefont
  {Baibhav}}, \bibinfo {author} {\bibfnamefont {E.}~\bibnamefont {Berti}},
  \bibinfo {author} {\bibfnamefont {V.}~\bibnamefont {Cardoso}}, \bibinfo
  {author} {\bibfnamefont {G.}~\bibnamefont {Carullo}}, \bibinfo {author}
  {\bibfnamefont {R.}~\bibnamefont {Cotesta}}, \bibinfo {author} {\bibfnamefont
  {W.}~\bibnamefont {Del~Pozzo}}, \bibinfo {author} {\bibfnamefont
  {F.}~\bibnamefont {Duque}}, \bibinfo {author} {\bibfnamefont
  {T.}~\bibnamefont {Helfer}}, \bibinfo {author} {\bibfnamefont
  {E.}~\bibnamefont {Shukla}},\ and\ \bibinfo {author} {\bibfnamefont {K.~W.}\
  \bibnamefont {Wong}},\ }\bibfield  {title} {\bibinfo {title} {Nonlinear
  effects in black hole ringdown},\ }\bibfield  {journal} {\bibinfo  {journal}
  {Physical Review Letters}\ }\textbf {\bibinfo {volume} {130}},\ \href
  {https://doi.org/10.1103/physrevlett.130.081401}
  {10.1103/physrevlett.130.081401} (\bibinfo {year} {2023})\BibitemShut
  {NoStop}%
\bibitem [{\citenamefont {{Pook-Kolb}}\ \emph
  {et~al.}(2020{\natexlab{a}})\citenamefont {{Pook-Kolb}}, \citenamefont
  {{Birnholtz}}, \citenamefont {{Jaramillo}}, \citenamefont {{Krishnan}},\ and\
  \citenamefont {{Schnetter}}}]{Pook-Kolb2020}%
  \BibitemOpen
  \bibfield  {author} {\bibinfo {author} {\bibfnamefont {D.}~\bibnamefont
  {{Pook-Kolb}}}, \bibinfo {author} {\bibfnamefont {O.}~\bibnamefont
  {{Birnholtz}}}, \bibinfo {author} {\bibfnamefont {J.~L.}\ \bibnamefont
  {{Jaramillo}}}, \bibinfo {author} {\bibfnamefont {B.}~\bibnamefont
  {{Krishnan}}},\ and\ \bibinfo {author} {\bibfnamefont {E.}~\bibnamefont
  {{Schnetter}}},\ }\bibfield  {title} {\bibinfo {title} {{Horizons in a binary
  black hole merger II: Fluxes, multipole moments and stability}},\ }\href
  {https://doi.org/10.48550/arXiv.2006.03940} {\bibfield  {journal} {\bibinfo
  {journal} {arXiv e-prints}\ ,\ \bibinfo {eid} {arXiv:2006.03940}} (\bibinfo
  {year} {2020}{\natexlab{a}})},\ \Eprint {https://arxiv.org/abs/2006.03940}
  {arXiv:2006.03940 [gr-qc]} \BibitemShut {NoStop}%
\bibitem [{\citenamefont {{London}}\ \emph {et~al.}(2014)\citenamefont
  {{London}}, \citenamefont {{Shoemaker}},\ and\ \citenamefont
  {{Healy}}}]{London_2014}%
  \BibitemOpen
  \bibfield  {author} {\bibinfo {author} {\bibfnamefont {L.}~\bibnamefont
  {{London}}}, \bibinfo {author} {\bibfnamefont {D.}~\bibnamefont
  {{Shoemaker}}},\ and\ \bibinfo {author} {\bibfnamefont {J.}~\bibnamefont
  {{Healy}}},\ }\bibfield  {title} {\bibinfo {title} {{Modeling ringdown:
  Beyond the fundamental quasinormal modes}},\ }\href
  {https://doi.org/10.1103/PhysRevD.90.124032} {\bibfield  {journal} {\bibinfo
  {journal} {\prd}\ }\textbf {\bibinfo {volume} {90}},\ \bibinfo {eid} {124032}
  (\bibinfo {year} {2014})},\ \Eprint {https://arxiv.org/abs/1404.3197}
  {arXiv:1404.3197 [gr-qc]} \BibitemShut {NoStop}%
\bibitem [{\citenamefont {London}(2020)}]{London_2020}%
  \BibitemOpen
  \bibfield  {author} {\bibinfo {author} {\bibfnamefont {L.}~\bibnamefont
  {London}},\ }\bibfield  {title} {\bibinfo {title} {Modeling ringdown. ii.
  aligned-spin binary black holes, implications for data analysis and
  fundamental theory},\ }\bibfield  {journal} {\bibinfo  {journal} {Physical
  Review D}\ }\textbf {\bibinfo {volume} {102}},\ \href
  {https://doi.org/10.1103/physrevd.102.084052} {10.1103/physrevd.102.084052}
  (\bibinfo {year} {2020})\BibitemShut {NoStop}%
\bibitem [{\citenamefont {{Jim{\'e}nez Forteza}}\ \emph
  {et~al.}(2020)\citenamefont {{Jim{\'e}nez Forteza}}, \citenamefont
  {{Bhagwat}}, \citenamefont {{Pani}},\ and\ \citenamefont
  {{Ferrari}}}]{JimenezForteza2020}%
  \BibitemOpen
  \bibfield  {author} {\bibinfo {author} {\bibfnamefont {X.}~\bibnamefont
  {{Jim{\'e}nez Forteza}}}, \bibinfo {author} {\bibfnamefont {S.}~\bibnamefont
  {{Bhagwat}}}, \bibinfo {author} {\bibfnamefont {P.}~\bibnamefont {{Pani}}},\
  and\ \bibinfo {author} {\bibfnamefont {V.}~\bibnamefont {{Ferrari}}},\
  }\bibfield  {title} {\bibinfo {title} {{Spectroscopy of binary black hole
  ringdown using overtones and angular modes}},\ }\href
  {https://doi.org/10.1103/PhysRevD.102.044053} {\bibfield  {journal} {\bibinfo
   {journal} {\prd}\ }\textbf {\bibinfo {volume} {102}},\ \bibinfo {eid}
  {044053} (\bibinfo {year} {2020})},\ \Eprint
  {https://arxiv.org/abs/2005.03260} {arXiv:2005.03260 [gr-qc]} \BibitemShut
  {NoStop}%
\bibitem [{\citenamefont {{Zhu}}\ \emph {et~al.}(2023)\citenamefont {{Zhu}},
  \citenamefont {{Ripley}}, \citenamefont {{C{\'a}rdenas-Avenda{\~n}o}},\ and\
  \citenamefont {{Pretorius}}}]{Zhu_2023}%
  \BibitemOpen
  \bibfield  {author} {\bibinfo {author} {\bibfnamefont {H.}~\bibnamefont
  {{Zhu}}}, \bibinfo {author} {\bibfnamefont {J.~L.}\ \bibnamefont {{Ripley}}},
  \bibinfo {author} {\bibfnamefont {A.}~\bibnamefont
  {{C{\'a}rdenas-Avenda{\~n}o}}},\ and\ \bibinfo {author} {\bibfnamefont
  {F.}~\bibnamefont {{Pretorius}}},\ }\bibfield  {title} {\bibinfo {title}
  {{Challenges in Quasinormal Mode Extraction: Perspectives from Numerical
  solutions to the Teukolsky Equation}},\ }\href
  {https://doi.org/10.48550/arXiv.2309.13204} {\bibfield  {journal} {\bibinfo
  {journal} {arXiv e-prints}\ ,\ \bibinfo {eid} {arXiv:2309.13204}} (\bibinfo
  {year} {2023})},\ \Eprint {https://arxiv.org/abs/2309.13204}
  {arXiv:2309.13204 [gr-qc]} \BibitemShut {NoStop}%
\bibitem [{\citenamefont {{Nee}}\ \emph {et~al.}(2023)\citenamefont {{Nee}},
  \citenamefont {{V{\"o}lkel}},\ and\ \citenamefont {{Pfeiffer}}}]{Nee_2023}%
  \BibitemOpen
  \bibfield  {author} {\bibinfo {author} {\bibfnamefont {P.~J.}\ \bibnamefont
  {{Nee}}}, \bibinfo {author} {\bibfnamefont {S.~H.}\ \bibnamefont
  {{V{\"o}lkel}}},\ and\ \bibinfo {author} {\bibfnamefont {H.~P.}\ \bibnamefont
  {{Pfeiffer}}},\ }\bibfield  {title} {\bibinfo {title} {{Role of black hole
  quasinormal mode overtones for ringdown analysis}},\ }\href
  {https://doi.org/10.1103/PhysRevD.108.044032} {\bibfield  {journal} {\bibinfo
   {journal} {\prd}\ }\textbf {\bibinfo {volume} {108}},\ \bibinfo {eid}
  {044032} (\bibinfo {year} {2023})},\ \Eprint
  {https://arxiv.org/abs/2302.06634} {arXiv:2302.06634 [gr-qc]} \BibitemShut
  {NoStop}%
\bibitem [{\citenamefont {{Ho-Yeuk Cheung}}\ \emph {et~al.}(2023)\citenamefont
  {{Ho-Yeuk Cheung}}, \citenamefont {{Berti}}, \citenamefont {{Baibhav}},\ and\
  \citenamefont {{Cotesta}}}]{Cheung_2023}%
  \BibitemOpen
  \bibfield  {author} {\bibinfo {author} {\bibfnamefont {M.}~\bibnamefont
  {{Ho-Yeuk Cheung}}}, \bibinfo {author} {\bibfnamefont {E.}~\bibnamefont
  {{Berti}}}, \bibinfo {author} {\bibfnamefont {V.}~\bibnamefont {{Baibhav}}},\
  and\ \bibinfo {author} {\bibfnamefont {R.}~\bibnamefont {{Cotesta}}},\
  }\bibfield  {title} {\bibinfo {title} {{Extracting linear and nonlinear
  quasinormal modes from black hole merger simulations}},\ }\href
  {https://doi.org/10.48550/arXiv.2310.04489} {\bibfield  {journal} {\bibinfo
  {journal} {arXiv e-prints}\ ,\ \bibinfo {eid} {arXiv:2310.04489}} (\bibinfo
  {year} {2023})},\ \Eprint {https://arxiv.org/abs/2310.04489}
  {arXiv:2310.04489 [gr-qc]} \BibitemShut {NoStop}%
\bibitem [{\citenamefont {{Redondo-Yuste}}\ \emph
  {et~al.}(2023{\natexlab{a}})\citenamefont {{Redondo-Yuste}}, \citenamefont
  {{Carullo}}, \citenamefont {{Ripley}}, \citenamefont {{Berti}},\ and\
  \citenamefont {{Cardoso}}}]{Redondo-Yuste2023b}%
  \BibitemOpen
  \bibfield  {author} {\bibinfo {author} {\bibfnamefont {J.}~\bibnamefont
  {{Redondo-Yuste}}}, \bibinfo {author} {\bibfnamefont {G.}~\bibnamefont
  {{Carullo}}}, \bibinfo {author} {\bibfnamefont {J.~L.}\ \bibnamefont
  {{Ripley}}}, \bibinfo {author} {\bibfnamefont {E.}~\bibnamefont {{Berti}}},\
  and\ \bibinfo {author} {\bibfnamefont {V.}~\bibnamefont {{Cardoso}}},\
  }\bibfield  {title} {\bibinfo {title} {{Spin dependence of black hole
  ringdown nonlinearities}},\ }\href
  {https://doi.org/10.48550/arXiv.2308.14796} {\bibfield  {journal} {\bibinfo
  {journal} {arXiv e-prints}\ ,\ \bibinfo {eid} {arXiv:2308.14796}} (\bibinfo
  {year} {2023}{\natexlab{a}})},\ \Eprint {https://arxiv.org/abs/2308.14796}
  {arXiv:2308.14796 [gr-qc]} \BibitemShut {NoStop}%
\bibitem [{\citenamefont {Carullo}\ \emph {et~al.}(2019)\citenamefont
  {Carullo}, \citenamefont {Pozzo},\ and\ \citenamefont
  {Veitch}}]{Carullo_2019}%
  \BibitemOpen
  \bibfield  {author} {\bibinfo {author} {\bibfnamefont {G.}~\bibnamefont
  {Carullo}}, \bibinfo {author} {\bibfnamefont {W.~D.}\ \bibnamefont {Pozzo}},\
  and\ \bibinfo {author} {\bibfnamefont {J.}~\bibnamefont {Veitch}},\
  }\bibfield  {title} {\bibinfo {title} {Observational black hole spectroscopy:
  A time-domain multimode analysis of {GW}150914},\ }\bibfield  {journal}
  {\bibinfo  {journal} {Physical Review D}\ }\textbf {\bibinfo {volume} {99}},\
  \href {https://doi.org/10.1103/physrevd.99.123029}
  {10.1103/physrevd.99.123029} (\bibinfo {year} {2019})\BibitemShut {NoStop}%
\bibitem [{\citenamefont {{Wang}}\ \emph {et~al.}(2023)\citenamefont {{Wang}},
  \citenamefont {{Capano}}, \citenamefont {{Abedi}}, \citenamefont {{Kastha}},
  \citenamefont {{Krishnan}}, \citenamefont {{Nielsen}}, \citenamefont
  {{Nitz}},\ and\ \citenamefont {{Westerweck}}}]{Wang_2023b}%
  \BibitemOpen
  \bibfield  {author} {\bibinfo {author} {\bibfnamefont {Y.-F.}\ \bibnamefont
  {{Wang}}}, \bibinfo {author} {\bibfnamefont {C.~D.}\ \bibnamefont
  {{Capano}}}, \bibinfo {author} {\bibfnamefont {J.}~\bibnamefont {{Abedi}}},
  \bibinfo {author} {\bibfnamefont {S.}~\bibnamefont {{Kastha}}}, \bibinfo
  {author} {\bibfnamefont {B.}~\bibnamefont {{Krishnan}}}, \bibinfo {author}
  {\bibfnamefont {A.~B.}\ \bibnamefont {{Nielsen}}}, \bibinfo {author}
  {\bibfnamefont {A.~H.}\ \bibnamefont {{Nitz}}},\ and\ \bibinfo {author}
  {\bibfnamefont {J.}~\bibnamefont {{Westerweck}}},\ }\bibfield  {title}
  {\bibinfo {title} {{A frequency-domain perspective on GW150914 ringdown
  overtone}},\ }\href {https://doi.org/10.48550/arXiv.2310.19645} {\bibfield
  {journal} {\bibinfo  {journal} {arXiv e-prints}\ ,\ \bibinfo {eid}
  {arXiv:2310.19645}} (\bibinfo {year} {2023})},\ \Eprint
  {https://arxiv.org/abs/2310.19645} {arXiv:2310.19645 [gr-qc]} \BibitemShut
  {NoStop}%
\bibitem [{\citenamefont {{Ma}}\ \emph {et~al.}(2023)\citenamefont {{Ma}},
  \citenamefont {{Sun}},\ and\ \citenamefont {{Chen}}}]{Ma_2023}%
  \BibitemOpen
  \bibfield  {author} {\bibinfo {author} {\bibfnamefont {S.}~\bibnamefont
  {{Ma}}}, \bibinfo {author} {\bibfnamefont {L.}~\bibnamefont {{Sun}}},\ and\
  \bibinfo {author} {\bibfnamefont {Y.}~\bibnamefont {{Chen}}},\ }\bibfield
  {title} {\bibinfo {title} {{Using rational filters to uncover the first
  ringdown overtone in GW150914}},\ }\href
  {https://doi.org/10.1103/PhysRevD.107.084010} {\bibfield  {journal} {\bibinfo
   {journal} {\prd}\ }\textbf {\bibinfo {volume} {107}},\ \bibinfo {eid}
  {084010} (\bibinfo {year} {2023})},\ \Eprint
  {https://arxiv.org/abs/2301.06639} {arXiv:2301.06639 [gr-qc]} \BibitemShut
  {NoStop}%
\bibitem [{\citenamefont {{Finch}}\ and\ \citenamefont
  {{Moore}}(2022)}]{Finch_2022}%
  \BibitemOpen
  \bibfield  {author} {\bibinfo {author} {\bibfnamefont {E.}~\bibnamefont
  {{Finch}}}\ and\ \bibinfo {author} {\bibfnamefont {C.~J.}\ \bibnamefont
  {{Moore}}},\ }\bibfield  {title} {\bibinfo {title} {{Searching for a ringdown
  overtone in GW150914}},\ }\href {https://doi.org/10.1103/PhysRevD.106.043005}
  {\bibfield  {journal} {\bibinfo  {journal} {\prd}\ }\textbf {\bibinfo
  {volume} {106}},\ \bibinfo {eid} {043005} (\bibinfo {year} {2022})},\ \Eprint
  {https://arxiv.org/abs/2205.07809} {arXiv:2205.07809 [gr-qc]} \BibitemShut
  {NoStop}%
\bibitem [{\citenamefont {{Wang}}\ and\ \citenamefont
  {{Shao}}(2023)}]{Wang_2023}%
  \BibitemOpen
  \bibfield  {author} {\bibinfo {author} {\bibfnamefont {H.-T.}\ \bibnamefont
  {{Wang}}}\ and\ \bibinfo {author} {\bibfnamefont {L.}~\bibnamefont
  {{Shao}}},\ }\bibfield  {title} {\bibinfo {title} {{Effect of Noise
  Estimation in Time-Domain Ringdown Analysis: A Case Study with GW150914}},\
  }\href@noop {} {\bibfield  {journal} {\bibinfo  {journal} {arXiv e-prints}\
  ,\ \bibinfo {eid} {arXiv:2311.13300}} (\bibinfo {year} {2023})},\ \Eprint
  {https://arxiv.org/abs/2311.13300} {arXiv:2311.13300 [gr-qc]} \BibitemShut
  {NoStop}%
\bibitem [{\citenamefont {{Crisostomi}}\ \emph {et~al.}(2023)\citenamefont
  {{Crisostomi}}, \citenamefont {{Dey}}, \citenamefont {{Barausse}},\ and\
  \citenamefont {{Trotta}}}]{Crisostomi2023}%
  \BibitemOpen
  \bibfield  {author} {\bibinfo {author} {\bibfnamefont {M.}~\bibnamefont
  {{Crisostomi}}}, \bibinfo {author} {\bibfnamefont {K.}~\bibnamefont {{Dey}}},
  \bibinfo {author} {\bibfnamefont {E.}~\bibnamefont {{Barausse}}},\ and\
  \bibinfo {author} {\bibfnamefont {R.}~\bibnamefont {{Trotta}}},\ }\bibfield
  {title} {\bibinfo {title} {{Neural posterior estimation with guaranteed exact
  coverage: The ringdown of GW150914}},\ }\href
  {https://doi.org/10.1103/PhysRevD.108.044029} {\bibfield  {journal} {\bibinfo
   {journal} {\prd}\ }\textbf {\bibinfo {volume} {108}},\ \bibinfo {eid}
  {044029} (\bibinfo {year} {2023})},\ \Eprint
  {https://arxiv.org/abs/2305.18528} {arXiv:2305.18528 [gr-qc]} \BibitemShut
  {NoStop}%
\bibitem [{\citenamefont {{Correia}}\ \emph {et~al.}(2023)\citenamefont
  {{Correia}}, \citenamefont {{Wang}},\ and\ \citenamefont
  {{Capano}}}]{Correia_2023}%
  \BibitemOpen
  \bibfield  {author} {\bibinfo {author} {\bibfnamefont {A.}~\bibnamefont
  {{Correia}}}, \bibinfo {author} {\bibfnamefont {Y.-F.}\ \bibnamefont
  {{Wang}}},\ and\ \bibinfo {author} {\bibfnamefont {C.~D.}\ \bibnamefont
  {{Capano}}},\ }\bibfield  {title} {\bibinfo {title} {{Low evidence for
  ringdown overtone in GW150914 when marginalizing over time and sky location
  uncertainty}},\ }\href {https://doi.org/10.48550/arXiv.2312.14118} {\bibfield
   {journal} {\bibinfo  {journal} {arXiv e-prints}\ ,\ \bibinfo {eid}
  {arXiv:2312.14118}} (\bibinfo {year} {2023})},\ \Eprint
  {https://arxiv.org/abs/2312.14118} {arXiv:2312.14118 [gr-qc]} \BibitemShut
  {NoStop}%
\bibitem [{\citenamefont {Isi}\ and\ \citenamefont {Farr}(2023)}]{Isi:2023nif}%
  \BibitemOpen
  \bibfield  {author} {\bibinfo {author} {\bibfnamefont {M.}~\bibnamefont
  {Isi}}\ and\ \bibinfo {author} {\bibfnamefont {W.~M.}\ \bibnamefont {Farr}},\
  }\bibfield  {title} {\bibinfo {title} {{Comment on
  \textquotedblleft{}Analysis of Ringdown Overtones in
  GW150914\textquotedblright{}}},\ }\href
  {https://doi.org/10.1103/PhysRevLett.131.169001} {\bibfield  {journal}
  {\bibinfo  {journal} {Phys. Rev. Lett.}\ }\textbf {\bibinfo {volume} {131}},\
  \bibinfo {pages} {169001} (\bibinfo {year} {2023})},\ \Eprint
  {https://arxiv.org/abs/2310.13869} {arXiv:2310.13869 [astro-ph.HE]}
  \BibitemShut {NoStop}%
\bibitem [{\citenamefont {Carullo}\ \emph {et~al.}(2023)\citenamefont
  {Carullo}, \citenamefont {Cotesta}, \citenamefont {Berti},\ and\
  \citenamefont {Cardoso}}]{Carullo:2023gtf}%
  \BibitemOpen
  \bibfield  {author} {\bibinfo {author} {\bibfnamefont {G.}~\bibnamefont
  {Carullo}}, \bibinfo {author} {\bibfnamefont {R.}~\bibnamefont {Cotesta}},
  \bibinfo {author} {\bibfnamefont {E.}~\bibnamefont {Berti}},\ and\ \bibinfo
  {author} {\bibfnamefont {V.}~\bibnamefont {Cardoso}},\ }\bibfield  {title}
  {\bibinfo {title} {{Reply to Comment on ''Analysis of Ringdown Overtones in
  GW150914''}},\ }\href {https://doi.org/10.1103/PhysRevLett.131.169002}
  {\bibfield  {journal} {\bibinfo  {journal} {Phys. Rev. Lett.}\ }\textbf
  {\bibinfo {volume} {131}},\ \bibinfo {pages} {169002} (\bibinfo {year}
  {2023})},\ \Eprint {https://arxiv.org/abs/2310.20625} {arXiv:2310.20625
  [gr-qc]} \BibitemShut {NoStop}%
\bibitem [{\citenamefont {{Berti}}\ \emph {et~al.}(2009)\citenamefont
  {{Berti}}, \citenamefont {{Cardoso}},\ and\ \citenamefont
  {{Starinets}}}]{Berti_2009}%
  \BibitemOpen
  \bibfield  {author} {\bibinfo {author} {\bibfnamefont {E.}~\bibnamefont
  {{Berti}}}, \bibinfo {author} {\bibfnamefont {V.}~\bibnamefont {{Cardoso}}},\
  and\ \bibinfo {author} {\bibfnamefont {A.~O.}\ \bibnamefont {{Starinets}}},\
  }\bibfield  {title} {\bibinfo {title} {{TOPICAL REVIEW: Quasinormal modes of
  black holes and black branes}},\ }\href
  {https://doi.org/10.1088/0264-9381/26/16/163001} {\bibfield  {journal}
  {\bibinfo  {journal} {Classical and Quantum Gravity}\ }\textbf {\bibinfo
  {volume} {26}},\ \bibinfo {eid} {163001} (\bibinfo {year} {2009})},\ \Eprint
  {https://arxiv.org/abs/0905.2975} {arXiv:0905.2975 [gr-qc]} \BibitemShut
  {NoStop}%
\bibitem [{\citenamefont {{Berti}}(2023)}]{berti_website}%
  \BibitemOpen
  \bibfield  {author} {\bibinfo {author} {\bibfnamefont {E.}~\bibnamefont
  {{Berti}}},\ }\href {https://pages.jh.edu/eberti2/ringdown/} {\bibinfo
  {title} {Ringdown}} (\bibinfo {year} {2023}),\ \bibinfo {note} {[Online;
  accessed 30-October-2023]}\BibitemShut {NoStop}%
\bibitem [{\citenamefont {{Teukolsky}}(1972)}]{Teukolsky_1972}%
  \BibitemOpen
  \bibfield  {author} {\bibinfo {author} {\bibfnamefont {S.~A.}\ \bibnamefont
  {{Teukolsky}}},\ }\bibfield  {title} {\bibinfo {title} {{Rotating Black
  Holes: Separable Wave Equations for Gravitational and Electromagnetic
  Perturbations}},\ }\href {https://doi.org/10.1103/PhysRevLett.29.1114}
  {\bibfield  {journal} {\bibinfo  {journal} {\prl}\ }\textbf {\bibinfo
  {volume} {29}},\ \bibinfo {pages} {1114} (\bibinfo {year}
  {1972})}\BibitemShut {NoStop}%
\bibitem [{\citenamefont {{Mrou{\'e}}}\ \emph {et~al.}(2013)\citenamefont
  {{Mrou{\'e}}}, \citenamefont {{Scheel}}, \citenamefont {{Szil{\'a}gyi}},
  \citenamefont {{Pfeiffer}}, \citenamefont {{Boyle}}, \citenamefont
  {{Hemberger}}, \citenamefont {{Kidder}}, \citenamefont {{Lovelace}},
  \citenamefont {{Ossokine}}, \citenamefont {{Taylor}}, \citenamefont
  {{Zengino{\u{g}}lu}}, \citenamefont {{Buchman}}, \citenamefont {{Chu}},
  \citenamefont {{Foley}}, \citenamefont {{Giesler}}, \citenamefont {{Owen}},\
  and\ \citenamefont {{Teukolsky}}}]{SXS_2013}%
  \BibitemOpen
  \bibfield  {author} {\bibinfo {author} {\bibfnamefont {A.~H.}\ \bibnamefont
  {{Mrou{\'e}}}}, \bibinfo {author} {\bibfnamefont {M.~A.}\ \bibnamefont
  {{Scheel}}}, \bibinfo {author} {\bibfnamefont {B.}~\bibnamefont
  {{Szil{\'a}gyi}}}, \bibinfo {author} {\bibfnamefont {H.~P.}\ \bibnamefont
  {{Pfeiffer}}}, \bibinfo {author} {\bibfnamefont {M.}~\bibnamefont {{Boyle}}},
  \bibinfo {author} {\bibfnamefont {D.~A.}\ \bibnamefont {{Hemberger}}},
  \bibinfo {author} {\bibfnamefont {L.~E.}\ \bibnamefont {{Kidder}}}, \bibinfo
  {author} {\bibfnamefont {G.}~\bibnamefont {{Lovelace}}}, \bibinfo {author}
  {\bibfnamefont {S.}~\bibnamefont {{Ossokine}}}, \bibinfo {author}
  {\bibfnamefont {N.~W.}\ \bibnamefont {{Taylor}}}, \bibinfo {author}
  {\bibfnamefont {A.}~\bibnamefont {{Zengino{\u{g}}lu}}}, \bibinfo {author}
  {\bibfnamefont {L.~T.}\ \bibnamefont {{Buchman}}}, \bibinfo {author}
  {\bibfnamefont {T.}~\bibnamefont {{Chu}}}, \bibinfo {author} {\bibfnamefont
  {E.}~\bibnamefont {{Foley}}}, \bibinfo {author} {\bibfnamefont
  {M.}~\bibnamefont {{Giesler}}}, \bibinfo {author} {\bibfnamefont
  {R.}~\bibnamefont {{Owen}}},\ and\ \bibinfo {author} {\bibfnamefont {S.~A.}\
  \bibnamefont {{Teukolsky}}},\ }\bibfield  {title} {\bibinfo {title} {{Catalog
  of 174 Binary Black Hole Simulations for Gravitational Wave Astronomy}},\
  }\href {https://doi.org/10.1103/PhysRevLett.111.241104} {\bibfield  {journal}
  {\bibinfo  {journal} {\prl}\ }\textbf {\bibinfo {volume} {111}},\ \bibinfo
  {eid} {241104} (\bibinfo {year} {2013})},\ \Eprint
  {https://arxiv.org/abs/1304.6077} {arXiv:1304.6077 [gr-qc]} \BibitemShut
  {NoStop}%
\bibitem [{\citenamefont {{Boyle}}\ \emph {et~al.}(2019)\citenamefont
  {{Boyle}}, \citenamefont {{Hemberger}}, \citenamefont {{Iozzo}},
  \citenamefont {{Lovelace}}, \citenamefont {{Ossokine}}, \citenamefont
  {{Pfeiffer}}, \citenamefont {{Scheel}}, \citenamefont {{Stein}},
  \citenamefont {{Woodford}}, \citenamefont {{Zimmerman}}, \citenamefont
  {{Afshari}}, \citenamefont {{Barkett}}, \citenamefont {{Blackman}},
  \citenamefont {{Chatziioannou}}, \citenamefont {{Chu}}, \citenamefont
  {{Demos}}, \citenamefont {{Deppe}}, \citenamefont {{Field}}, \citenamefont
  {{Fischer}}, \citenamefont {{Foley}}, \citenamefont {{Fong}}, \citenamefont
  {{Garcia}}, \citenamefont {{Giesler}}, \citenamefont {{Hebert}},
  \citenamefont {{Hinder}}, \citenamefont {{Katebi}}, \citenamefont {{Khan}},
  \citenamefont {{Kidder}}, \citenamefont {{Kumar}}, \citenamefont {{Kuper}},
  \citenamefont {{Lim}}, \citenamefont {{Okounkova}}, \citenamefont
  {{Ramirez}}, \citenamefont {{Rodriguez}}, \citenamefont {{R{\"u}ter}},
  \citenamefont {{Schmidt}}, \citenamefont {{Szilagyi}}, \citenamefont
  {{Teukolsky}}, \citenamefont {{Varma}},\ and\ \citenamefont
  {{Walker}}}]{SXS_2019}%
  \BibitemOpen
  \bibfield  {author} {\bibinfo {author} {\bibfnamefont {M.}~\bibnamefont
  {{Boyle}}}, \bibinfo {author} {\bibfnamefont {D.}~\bibnamefont
  {{Hemberger}}}, \bibinfo {author} {\bibfnamefont {D.~A.~B.}\ \bibnamefont
  {{Iozzo}}}, \bibinfo {author} {\bibfnamefont {G.}~\bibnamefont {{Lovelace}}},
  \bibinfo {author} {\bibfnamefont {S.}~\bibnamefont {{Ossokine}}}, \bibinfo
  {author} {\bibfnamefont {H.~P.}\ \bibnamefont {{Pfeiffer}}}, \bibinfo
  {author} {\bibfnamefont {M.~A.}\ \bibnamefont {{Scheel}}}, \bibinfo {author}
  {\bibfnamefont {L.~C.}\ \bibnamefont {{Stein}}}, \bibinfo {author}
  {\bibfnamefont {C.~J.}\ \bibnamefont {{Woodford}}}, \bibinfo {author}
  {\bibfnamefont {A.~B.}\ \bibnamefont {{Zimmerman}}}, \bibinfo {author}
  {\bibfnamefont {N.}~\bibnamefont {{Afshari}}}, \bibinfo {author}
  {\bibfnamefont {K.}~\bibnamefont {{Barkett}}}, \bibinfo {author}
  {\bibfnamefont {J.}~\bibnamefont {{Blackman}}}, \bibinfo {author}
  {\bibfnamefont {K.}~\bibnamefont {{Chatziioannou}}}, \bibinfo {author}
  {\bibfnamefont {T.}~\bibnamefont {{Chu}}}, \bibinfo {author} {\bibfnamefont
  {N.}~\bibnamefont {{Demos}}}, \bibinfo {author} {\bibfnamefont
  {N.}~\bibnamefont {{Deppe}}}, \bibinfo {author} {\bibfnamefont {S.~E.}\
  \bibnamefont {{Field}}}, \bibinfo {author} {\bibfnamefont {N.~L.}\
  \bibnamefont {{Fischer}}}, \bibinfo {author} {\bibfnamefont {E.}~\bibnamefont
  {{Foley}}}, \bibinfo {author} {\bibfnamefont {H.}~\bibnamefont {{Fong}}},
  \bibinfo {author} {\bibfnamefont {A.}~\bibnamefont {{Garcia}}}, \bibinfo
  {author} {\bibfnamefont {M.}~\bibnamefont {{Giesler}}}, \bibinfo {author}
  {\bibfnamefont {F.}~\bibnamefont {{Hebert}}}, \bibinfo {author}
  {\bibfnamefont {I.}~\bibnamefont {{Hinder}}}, \bibinfo {author}
  {\bibfnamefont {R.}~\bibnamefont {{Katebi}}}, \bibinfo {author}
  {\bibfnamefont {H.}~\bibnamefont {{Khan}}}, \bibinfo {author} {\bibfnamefont
  {L.~E.}\ \bibnamefont {{Kidder}}}, \bibinfo {author} {\bibfnamefont
  {P.}~\bibnamefont {{Kumar}}}, \bibinfo {author} {\bibfnamefont
  {K.}~\bibnamefont {{Kuper}}}, \bibinfo {author} {\bibfnamefont
  {H.}~\bibnamefont {{Lim}}}, \bibinfo {author} {\bibfnamefont
  {M.}~\bibnamefont {{Okounkova}}}, \bibinfo {author} {\bibfnamefont
  {T.}~\bibnamefont {{Ramirez}}}, \bibinfo {author} {\bibfnamefont
  {S.}~\bibnamefont {{Rodriguez}}}, \bibinfo {author} {\bibfnamefont {H.~R.}\
  \bibnamefont {{R{\"u}ter}}}, \bibinfo {author} {\bibfnamefont
  {P.}~\bibnamefont {{Schmidt}}}, \bibinfo {author} {\bibfnamefont
  {B.}~\bibnamefont {{Szilagyi}}}, \bibinfo {author} {\bibfnamefont {S.~A.}\
  \bibnamefont {{Teukolsky}}}, \bibinfo {author} {\bibfnamefont
  {V.}~\bibnamefont {{Varma}}},\ and\ \bibinfo {author} {\bibfnamefont
  {M.}~\bibnamefont {{Walker}}},\ }\bibfield  {title} {\bibinfo {title} {{The
  SXS collaboration catalog of binary black hole simulations}},\ }\href
  {https://doi.org/10.1088/1361-6382/ab34e2} {\bibfield  {journal} {\bibinfo
  {journal} {Classical and Quantum Gravity}\ }\textbf {\bibinfo {volume}
  {36}},\ \bibinfo {eid} {195006} (\bibinfo {year} {2019})},\ \Eprint
  {https://arxiv.org/abs/1904.04831} {arXiv:1904.04831 [gr-qc]} \BibitemShut
  {NoStop}%
\bibitem [{\citenamefont {{Mitman}}\ \emph {et~al.}(2020)\citenamefont
  {{Mitman}}, \citenamefont {{Moxon}}, \citenamefont {{Scheel}}, \citenamefont
  {{Teukolsky}}, \citenamefont {{Boyle}}, \citenamefont {{Deppe}},
  \citenamefont {{Kidder}},\ and\ \citenamefont {{Throwe}}}]{Mitman_2020}%
  \BibitemOpen
  \bibfield  {author} {\bibinfo {author} {\bibfnamefont {K.}~\bibnamefont
  {{Mitman}}}, \bibinfo {author} {\bibfnamefont {J.}~\bibnamefont {{Moxon}}},
  \bibinfo {author} {\bibfnamefont {M.~A.}\ \bibnamefont {{Scheel}}}, \bibinfo
  {author} {\bibfnamefont {S.~A.}\ \bibnamefont {{Teukolsky}}}, \bibinfo
  {author} {\bibfnamefont {M.}~\bibnamefont {{Boyle}}}, \bibinfo {author}
  {\bibfnamefont {N.}~\bibnamefont {{Deppe}}}, \bibinfo {author} {\bibfnamefont
  {L.~E.}\ \bibnamefont {{Kidder}}},\ and\ \bibinfo {author} {\bibfnamefont
  {W.}~\bibnamefont {{Throwe}}},\ }\bibfield  {title} {\bibinfo {title}
  {{Computation of displacement and spin gravitational memory in numerical
  relativity}},\ }\href {https://doi.org/10.1103/PhysRevD.102.104007}
  {\bibfield  {journal} {\bibinfo  {journal} {\prd}\ }\textbf {\bibinfo
  {volume} {102}},\ \bibinfo {eid} {104007} (\bibinfo {year} {2020})},\ \Eprint
  {https://arxiv.org/abs/2007.11562} {arXiv:2007.11562 [gr-qc]} \BibitemShut
  {NoStop}%
\bibitem [{\citenamefont {{Moxon}}\ \emph {et~al.}(2020)\citenamefont
  {{Moxon}}, \citenamefont {{Scheel}},\ and\ \citenamefont
  {{Teukolsky}}}]{Moxon_2020}%
  \BibitemOpen
  \bibfield  {author} {\bibinfo {author} {\bibfnamefont {J.}~\bibnamefont
  {{Moxon}}}, \bibinfo {author} {\bibfnamefont {M.~A.}\ \bibnamefont
  {{Scheel}}},\ and\ \bibinfo {author} {\bibfnamefont {S.~A.}\ \bibnamefont
  {{Teukolsky}}},\ }\bibfield  {title} {\bibinfo {title} {{Improved
  Cauchy-characteristic evolution system for high-precision numerical
  relativity waveforms}},\ }\href {https://doi.org/10.1103/PhysRevD.102.044052}
  {\bibfield  {journal} {\bibinfo  {journal} {\prd}\ }\textbf {\bibinfo
  {volume} {102}},\ \bibinfo {eid} {044052} (\bibinfo {year} {2020})},\ \Eprint
  {https://arxiv.org/abs/2007.01339} {arXiv:2007.01339 [gr-qc]} \BibitemShut
  {NoStop}%
\bibitem [{\citenamefont {{Moxon}}\ \emph {et~al.}(2021)\citenamefont
  {{Moxon}}, \citenamefont {{Scheel}}, \citenamefont {{Teukolsky}},
  \citenamefont {{Deppe}}, \citenamefont {{Fischer}}, \citenamefont
  {{H{\'e}bert}}, \citenamefont {{Kidder}},\ and\ \citenamefont
  {{Throwe}}}]{Moxon_2021}%
  \BibitemOpen
  \bibfield  {author} {\bibinfo {author} {\bibfnamefont {J.}~\bibnamefont
  {{Moxon}}}, \bibinfo {author} {\bibfnamefont {M.~A.}\ \bibnamefont
  {{Scheel}}}, \bibinfo {author} {\bibfnamefont {S.~A.}\ \bibnamefont
  {{Teukolsky}}}, \bibinfo {author} {\bibfnamefont {N.}~\bibnamefont
  {{Deppe}}}, \bibinfo {author} {\bibfnamefont {N.}~\bibnamefont {{Fischer}}},
  \bibinfo {author} {\bibfnamefont {F.}~\bibnamefont {{H{\'e}bert}}}, \bibinfo
  {author} {\bibfnamefont {L.~E.}\ \bibnamefont {{Kidder}}},\ and\ \bibinfo
  {author} {\bibfnamefont {W.}~\bibnamefont {{Throwe}}},\ }\bibfield  {title}
  {\bibinfo {title} {{The SpECTRE Cauchy-characteristic evolution system for
  rapid, precise waveform extraction}},\ }\href
  {https://doi.org/10.48550/arXiv.2110.08635} {\bibfield  {journal} {\bibinfo
  {journal} {arXiv e-prints}\ ,\ \bibinfo {eid} {arXiv:2110.08635}} (\bibinfo
  {year} {2021})},\ \Eprint {https://arxiv.org/abs/2110.08635}
  {arXiv:2110.08635 [gr-qc]} \BibitemShut {NoStop}%
\bibitem [{\citenamefont {Deppe}\ \emph {et~al.}(2023)\citenamefont {Deppe},
  \citenamefont {Throwe}, \citenamefont {Kidder}, \citenamefont {Vu},
  \citenamefont {Nelli}, \citenamefont {Armaza}, \citenamefont {Bonilla},
  \citenamefont {Hébert}, \citenamefont {Kim}, \citenamefont {Kumar},
  \citenamefont {Lovelace}, \citenamefont {Macedo}, \citenamefont {Moxon},
  \citenamefont {O'Shea}, \citenamefont {Pfeiffer}, \citenamefont {Scheel},
  \citenamefont {Teukolsky}, \citenamefont {Wittek}, \citenamefont
  {Anantpurkar}, \citenamefont {Anderson}, \citenamefont {Boyle}, \citenamefont
  {Carpenter}, \citenamefont {Ceja}, \citenamefont {Chaudhary}, \citenamefont
  {Corso}, \citenamefont {Foucart}, \citenamefont {Ghadiri}, \citenamefont
  {Giesler}, \citenamefont {Guo}, \citenamefont {Iozzo}, \citenamefont {Jones},
  \citenamefont {Lara}, \citenamefont {Legred}, \citenamefont {Li},
  \citenamefont {Ma}, \citenamefont {Melchor}, \citenamefont {Morales},
  \citenamefont {Most}, \citenamefont {Nee}, \citenamefont {Osorio},
  \citenamefont {Pajkos}, \citenamefont {Pannone}, \citenamefont {Ramirez},
  \citenamefont {Ring}, \citenamefont {Rüter}, \citenamefont {Sanchez},
  \citenamefont {Stein}, \citenamefont {Tellez}, \citenamefont {Thomas},
  \citenamefont {Vieira}, \citenamefont {Wlodarczyk}, \citenamefont {Wu},\ and\
  \citenamefont {Yoo}}]{SPECTRE}%
  \BibitemOpen
  \bibfield  {author} {\bibinfo {author} {\bibfnamefont {N.}~\bibnamefont
  {Deppe}}, \bibinfo {author} {\bibfnamefont {W.}~\bibnamefont {Throwe}},
  \bibinfo {author} {\bibfnamefont {L.~E.}\ \bibnamefont {Kidder}}, \bibinfo
  {author} {\bibfnamefont {N.~L.}\ \bibnamefont {Vu}}, \bibinfo {author}
  {\bibfnamefont {K.~C.}\ \bibnamefont {Nelli}}, \bibinfo {author}
  {\bibfnamefont {C.}~\bibnamefont {Armaza}}, \bibinfo {author} {\bibfnamefont
  {M.~S.}\ \bibnamefont {Bonilla}}, \bibinfo {author} {\bibfnamefont
  {F.}~\bibnamefont {Hébert}}, \bibinfo {author} {\bibfnamefont
  {Y.}~\bibnamefont {Kim}}, \bibinfo {author} {\bibfnamefont {P.}~\bibnamefont
  {Kumar}}, \bibinfo {author} {\bibfnamefont {G.}~\bibnamefont {Lovelace}},
  \bibinfo {author} {\bibfnamefont {A.}~\bibnamefont {Macedo}}, \bibinfo
  {author} {\bibfnamefont {J.}~\bibnamefont {Moxon}}, \bibinfo {author}
  {\bibfnamefont {E.}~\bibnamefont {O'Shea}}, \bibinfo {author} {\bibfnamefont
  {H.~P.}\ \bibnamefont {Pfeiffer}}, \bibinfo {author} {\bibfnamefont {M.~A.}\
  \bibnamefont {Scheel}}, \bibinfo {author} {\bibfnamefont {S.~A.}\
  \bibnamefont {Teukolsky}}, \bibinfo {author} {\bibfnamefont {N.~A.}\
  \bibnamefont {Wittek}}, \bibinfo {author} {\bibfnamefont {I.}~\bibnamefont
  {Anantpurkar}}, \bibinfo {author} {\bibfnamefont {C.}~\bibnamefont
  {Anderson}}, \bibinfo {author} {\bibfnamefont {M.}~\bibnamefont {Boyle}},
  \bibinfo {author} {\bibfnamefont {A.}~\bibnamefont {Carpenter}}, \bibinfo
  {author} {\bibfnamefont {A.}~\bibnamefont {Ceja}}, \bibinfo {author}
  {\bibfnamefont {H.}~\bibnamefont {Chaudhary}}, \bibinfo {author}
  {\bibfnamefont {N.}~\bibnamefont {Corso}}, \bibinfo {author} {\bibfnamefont
  {F.}~\bibnamefont {Foucart}}, \bibinfo {author} {\bibfnamefont
  {N.}~\bibnamefont {Ghadiri}}, \bibinfo {author} {\bibfnamefont
  {M.}~\bibnamefont {Giesler}}, \bibinfo {author} {\bibfnamefont {J.~S.}\
  \bibnamefont {Guo}}, \bibinfo {author} {\bibfnamefont {D.~A.~B.}\
  \bibnamefont {Iozzo}}, \bibinfo {author} {\bibfnamefont {K.~Z.}\ \bibnamefont
  {Jones}}, \bibinfo {author} {\bibfnamefont {G.}~\bibnamefont {Lara}},
  \bibinfo {author} {\bibfnamefont {I.}~\bibnamefont {Legred}}, \bibinfo
  {author} {\bibfnamefont {D.}~\bibnamefont {Li}}, \bibinfo {author}
  {\bibfnamefont {S.}~\bibnamefont {Ma}}, \bibinfo {author} {\bibfnamefont
  {D.}~\bibnamefont {Melchor}}, \bibinfo {author} {\bibfnamefont
  {M.}~\bibnamefont {Morales}}, \bibinfo {author} {\bibfnamefont {E.~R.}\
  \bibnamefont {Most}}, \bibinfo {author} {\bibfnamefont {P.~J.}\ \bibnamefont
  {Nee}}, \bibinfo {author} {\bibfnamefont {A.}~\bibnamefont {Osorio}},
  \bibinfo {author} {\bibfnamefont {M.~A.}\ \bibnamefont {Pajkos}}, \bibinfo
  {author} {\bibfnamefont {K.}~\bibnamefont {Pannone}}, \bibinfo {author}
  {\bibfnamefont {T.}~\bibnamefont {Ramirez}}, \bibinfo {author} {\bibfnamefont
  {N.}~\bibnamefont {Ring}}, \bibinfo {author} {\bibfnamefont {H.~R.}\
  \bibnamefont {Rüter}}, \bibinfo {author} {\bibfnamefont {J.}~\bibnamefont
  {Sanchez}}, \bibinfo {author} {\bibfnamefont {L.~C.}\ \bibnamefont {Stein}},
  \bibinfo {author} {\bibfnamefont {D.}~\bibnamefont {Tellez}}, \bibinfo
  {author} {\bibfnamefont {S.}~\bibnamefont {Thomas}}, \bibinfo {author}
  {\bibfnamefont {D.}~\bibnamefont {Vieira}}, \bibinfo {author} {\bibfnamefont
  {T.}~\bibnamefont {Wlodarczyk}}, \bibinfo {author} {\bibfnamefont
  {D.}~\bibnamefont {Wu}},\ and\ \bibinfo {author} {\bibfnamefont
  {J.}~\bibnamefont {Yoo}},\ }\href {https://doi.org/10.5281/zenodo.8431874}
  {\bibinfo {title} {Spectre}} (\bibinfo {year} {2023})\BibitemShut {NoStop}%
\bibitem [{\citenamefont {Mitman}\ \emph {et~al.}(2021)\citenamefont {Mitman},
  \citenamefont {Khera}, \citenamefont {Iozzo}, \citenamefont {Stein},
  \citenamefont {Boyle}, \citenamefont {Deppe}, \citenamefont {Kidder},
  \citenamefont {Moxon}, \citenamefont {Pfeiffer}, \citenamefont {Scheel},
  \citenamefont {Teukolsky},\ and\ \citenamefont {Throwe}}]{Mitman_2021}%
  \BibitemOpen
  \bibfield  {author} {\bibinfo {author} {\bibfnamefont {K.}~\bibnamefont
  {Mitman}}, \bibinfo {author} {\bibfnamefont {N.}~\bibnamefont {Khera}},
  \bibinfo {author} {\bibfnamefont {D.~A.~B.}\ \bibnamefont {Iozzo}}, \bibinfo
  {author} {\bibfnamefont {L.~C.}\ \bibnamefont {Stein}}, \bibinfo {author}
  {\bibfnamefont {M.}~\bibnamefont {Boyle}}, \bibinfo {author} {\bibfnamefont
  {N.}~\bibnamefont {Deppe}}, \bibinfo {author} {\bibfnamefont {L.~E.}\
  \bibnamefont {Kidder}}, \bibinfo {author} {\bibfnamefont {J.}~\bibnamefont
  {Moxon}}, \bibinfo {author} {\bibfnamefont {H.~P.}\ \bibnamefont {Pfeiffer}},
  \bibinfo {author} {\bibfnamefont {M.~A.}\ \bibnamefont {Scheel}}, \bibinfo
  {author} {\bibfnamefont {S.~A.}\ \bibnamefont {Teukolsky}},\ and\ \bibinfo
  {author} {\bibfnamefont {W.}~\bibnamefont {Throwe}},\ }\bibfield  {title}
  {\bibinfo {title} {Fixing the bms frame of numerical relativity waveforms},\
  }\bibfield  {journal} {\bibinfo  {journal} {Physical Review D}\ }\textbf
  {\bibinfo {volume} {104}},\ \href
  {https://doi.org/10.1103/physrevd.104.024051} {10.1103/physrevd.104.024051}
  (\bibinfo {year} {2021})\BibitemShut {NoStop}%
\bibitem [{\citenamefont {Mitman}\ \emph {et~al.}(2022)\citenamefont {Mitman},
  \citenamefont {Stein}, \citenamefont {Boyle}, \citenamefont {Deppe},
  \citenamefont {Hébert}, \citenamefont {Kidder}, \citenamefont {Moxon},
  \citenamefont {Scheel}, \citenamefont {Teukolsky}, \citenamefont {Throwe},\
  and\ \citenamefont {Vu}}]{Mitman_2022}%
  \BibitemOpen
  \bibfield  {author} {\bibinfo {author} {\bibfnamefont {K.}~\bibnamefont
  {Mitman}}, \bibinfo {author} {\bibfnamefont {L.~C.}\ \bibnamefont {Stein}},
  \bibinfo {author} {\bibfnamefont {M.}~\bibnamefont {Boyle}}, \bibinfo
  {author} {\bibfnamefont {N.}~\bibnamefont {Deppe}}, \bibinfo {author}
  {\bibfnamefont {F.}~\bibnamefont {Hébert}}, \bibinfo {author} {\bibfnamefont
  {L.~E.}\ \bibnamefont {Kidder}}, \bibinfo {author} {\bibfnamefont
  {J.}~\bibnamefont {Moxon}}, \bibinfo {author} {\bibfnamefont {M.~A.}\
  \bibnamefont {Scheel}}, \bibinfo {author} {\bibfnamefont {S.~A.}\
  \bibnamefont {Teukolsky}}, \bibinfo {author} {\bibfnamefont {W.}~\bibnamefont
  {Throwe}},\ and\ \bibinfo {author} {\bibfnamefont {N.~L.}\ \bibnamefont
  {Vu}},\ }\bibfield  {title} {\bibinfo {title} {Fixing the bms frame of
  numerical relativity waveforms with bms charges},\ }\bibfield  {journal}
  {\bibinfo  {journal} {Physical Review D}\ }\textbf {\bibinfo {volume}
  {106}},\ \href {https://doi.org/10.1103/physrevd.106.084029}
  {10.1103/physrevd.106.084029} (\bibinfo {year} {2022})\BibitemShut {NoStop}%
\bibitem [{\citenamefont {Boyle}(2013)}]{Boyle_2013}%
  \BibitemOpen
  \bibfield  {author} {\bibinfo {author} {\bibfnamefont {M.}~\bibnamefont
  {Boyle}},\ }\bibfield  {title} {\bibinfo {title} {Angular velocity of
  gravitational radiation from precessing binaries and the corotating frame},\
  }\bibfield  {journal} {\bibinfo  {journal} {Physical Review D}\ }\textbf
  {\bibinfo {volume} {87}},\ \href {https://doi.org/10.1103/physrevd.87.104006}
  {10.1103/physrevd.87.104006} (\bibinfo {year} {2013})\BibitemShut {NoStop}%
\bibitem [{\citenamefont {{Boyle}}\ \emph {et~al.}(2014)\citenamefont
  {{Boyle}}, \citenamefont {{Kidder}}, \citenamefont {{Ossokine}},\ and\
  \citenamefont {{Pfeiffer}}}]{Boyle_2014}%
  \BibitemOpen
  \bibfield  {author} {\bibinfo {author} {\bibfnamefont {M.}~\bibnamefont
  {{Boyle}}}, \bibinfo {author} {\bibfnamefont {L.~E.}\ \bibnamefont
  {{Kidder}}}, \bibinfo {author} {\bibfnamefont {S.}~\bibnamefont
  {{Ossokine}}},\ and\ \bibinfo {author} {\bibfnamefont {H.~P.}\ \bibnamefont
  {{Pfeiffer}}},\ }\bibfield  {title} {\bibinfo {title} {{Gravitational-wave
  modes from precessing black-hole binaries}},\ }\href
  {https://doi.org/10.48550/arXiv.1409.4431} {\bibfield  {journal} {\bibinfo
  {journal} {arXiv e-prints}\ ,\ \bibinfo {eid} {arXiv:1409.4431}} (\bibinfo
  {year} {2014})},\ \Eprint {https://arxiv.org/abs/1409.4431} {arXiv:1409.4431
  [gr-qc]} \BibitemShut {NoStop}%
\bibitem [{\citenamefont {Boyle}(2016)}]{Boyle_2016}%
  \BibitemOpen
  \bibfield  {author} {\bibinfo {author} {\bibfnamefont {M.}~\bibnamefont
  {Boyle}},\ }\bibfield  {title} {\bibinfo {title} {Transformations of
  asymptotic gravitational-wave data},\ }\bibfield  {journal} {\bibinfo
  {journal} {Physical Review D}\ }\textbf {\bibinfo {volume} {93}},\ \href
  {https://doi.org/10.1103/physrevd.93.084031} {10.1103/physrevd.93.084031}
  (\bibinfo {year} {2016})\BibitemShut {NoStop}%
\bibitem [{\citenamefont {Boyle}\ \emph {et~al.}(2023)\citenamefont {Boyle},
  \citenamefont {Iozzo}, \citenamefont {Stein}, \citenamefont {Khairnar},
  \citenamefont {Rüter}, \citenamefont {Scheel}, \citenamefont {Varma},\ and\
  \citenamefont {Mitman}}]{scri}%
  \BibitemOpen
  \bibfield  {author} {\bibinfo {author} {\bibfnamefont {M.}~\bibnamefont
  {Boyle}}, \bibinfo {author} {\bibfnamefont {D.}~\bibnamefont {Iozzo}},
  \bibinfo {author} {\bibfnamefont {L.}~\bibnamefont {Stein}}, \bibinfo
  {author} {\bibfnamefont {A.}~\bibnamefont {Khairnar}}, \bibinfo {author}
  {\bibfnamefont {H.}~\bibnamefont {Rüter}}, \bibinfo {author} {\bibfnamefont
  {M.}~\bibnamefont {Scheel}}, \bibinfo {author} {\bibfnamefont
  {V.}~\bibnamefont {Varma}},\ and\ \bibinfo {author} {\bibfnamefont
  {K.}~\bibnamefont {Mitman}},\ }\href
  {https://doi.org/10.5281/zenodo.10081312} {\bibinfo {title} {scri}} (\bibinfo
  {year} {2023})\BibitemShut {NoStop}%
\bibitem [{\citenamefont {{Abbott}}\ \emph
  {et~al.}(2016{\natexlab{b}})\citenamefont {{Abbott}} \emph
  {et~al.}}]{GW150914}%
  \BibitemOpen
  \bibfield  {author} {\bibinfo {author} {\bibfnamefont {B.~P.}\ \bibnamefont
  {{Abbott}}} \emph {et~al.},\ }\bibfield  {title} {\bibinfo {title}
  {{Observation of Gravitational Waves from a Binary Black Hole Merger}},\
  }\href {https://doi.org/10.1103/PhysRevLett.116.061102} {\bibfield  {journal}
  {\bibinfo  {journal} {\prl}\ }\textbf {\bibinfo {volume} {116}},\ \bibinfo
  {eid} {061102} (\bibinfo {year} {2016}{\natexlab{b}})},\ \Eprint
  {https://arxiv.org/abs/1602.03837} {arXiv:1602.03837 [gr-qc]} \BibitemShut
  {NoStop}%
\bibitem [{\citenamefont {{Foreman-Mackey}}\ \emph {et~al.}(2013)\citenamefont
  {{Foreman-Mackey}}, \citenamefont {{Hogg}}, \citenamefont {{Lang}},\ and\
  \citenamefont {{Goodman}}}]{emcee_2013}%
  \BibitemOpen
  \bibfield  {author} {\bibinfo {author} {\bibfnamefont {D.}~\bibnamefont
  {{Foreman-Mackey}}}, \bibinfo {author} {\bibfnamefont {D.~W.}\ \bibnamefont
  {{Hogg}}}, \bibinfo {author} {\bibfnamefont {D.}~\bibnamefont {{Lang}}},\
  and\ \bibinfo {author} {\bibfnamefont {J.}~\bibnamefont {{Goodman}}},\
  }\bibfield  {title} {\bibinfo {title} {{emcee: The MCMC Hammer}},\ }\href
  {https://doi.org/10.1086/670067} {\bibfield  {journal} {\bibinfo  {journal}
  {\pasp}\ }\textbf {\bibinfo {volume} {125}},\ \bibinfo {pages} {306}
  (\bibinfo {year} {2013})},\ \Eprint {https://arxiv.org/abs/1202.3665}
  {arXiv:1202.3665 [astro-ph.IM]} \BibitemShut {NoStop}%
\bibitem [{\citenamefont {{Pook-Kolb}}\ \emph
  {et~al.}(2020{\natexlab{b}})\citenamefont {{Pook-Kolb}}, \citenamefont
  {{Birnholtz}}, \citenamefont {{Jaramillo}}, \citenamefont {{Krishnan}},\ and\
  \citenamefont {{Schnetter}}}]{Pook-Kolb2020b}%
  \BibitemOpen
  \bibfield  {author} {\bibinfo {author} {\bibfnamefont {D.}~\bibnamefont
  {{Pook-Kolb}}}, \bibinfo {author} {\bibfnamefont {O.}~\bibnamefont
  {{Birnholtz}}}, \bibinfo {author} {\bibfnamefont {J.~L.}\ \bibnamefont
  {{Jaramillo}}}, \bibinfo {author} {\bibfnamefont {B.}~\bibnamefont
  {{Krishnan}}},\ and\ \bibinfo {author} {\bibfnamefont {E.}~\bibnamefont
  {{Schnetter}}},\ }\bibfield  {title} {\bibinfo {title} {{Horizons in a binary
  black hole merger II: Fluxes, multipole moments and stability}},\ }\href
  {https://doi.org/10.48550/arXiv.2006.03940} {\bibfield  {journal} {\bibinfo
  {journal} {arXiv e-prints}\ ,\ \bibinfo {eid} {arXiv:2006.03940}} (\bibinfo
  {year} {2020}{\natexlab{b}})},\ \Eprint {https://arxiv.org/abs/2006.03940}
  {arXiv:2006.03940 [gr-qc]} \BibitemShut {NoStop}%
\bibitem [{\citenamefont {{Sberna}}\ \emph {et~al.}(2022)\citenamefont
  {{Sberna}}, \citenamefont {{Bosch}}, \citenamefont {{East}}, \citenamefont
  {{Green}},\ and\ \citenamefont {{Lehner}}}]{Sberna2022}%
  \BibitemOpen
  \bibfield  {author} {\bibinfo {author} {\bibfnamefont {L.}~\bibnamefont
  {{Sberna}}}, \bibinfo {author} {\bibfnamefont {P.}~\bibnamefont {{Bosch}}},
  \bibinfo {author} {\bibfnamefont {W.~E.}\ \bibnamefont {{East}}}, \bibinfo
  {author} {\bibfnamefont {S.~R.}\ \bibnamefont {{Green}}},\ and\ \bibinfo
  {author} {\bibfnamefont {L.}~\bibnamefont {{Lehner}}},\ }\bibfield  {title}
  {\bibinfo {title} {{Nonlinear effects in the black hole ringdown:
  Absorption-induced mode excitation}},\ }\href
  {https://doi.org/10.1103/PhysRevD.105.064046} {\bibfield  {journal} {\bibinfo
   {journal} {\prd}\ }\textbf {\bibinfo {volume} {105}},\ \bibinfo {eid}
  {064046} (\bibinfo {year} {2022})},\ \Eprint
  {https://arxiv.org/abs/2112.11168} {arXiv:2112.11168 [gr-qc]} \BibitemShut
  {NoStop}%
\bibitem [{\citenamefont {{Redondo-Yuste}}\ \emph
  {et~al.}(2023{\natexlab{b}})\citenamefont {{Redondo-Yuste}}, \citenamefont
  {{Pere{\~n}iguez}},\ and\ \citenamefont {{Cardoso}}}]{Redondo-Yuste2023}%
  \BibitemOpen
  \bibfield  {author} {\bibinfo {author} {\bibfnamefont {J.}~\bibnamefont
  {{Redondo-Yuste}}}, \bibinfo {author} {\bibfnamefont {D.}~\bibnamefont
  {{Pere{\~n}iguez}}},\ and\ \bibinfo {author} {\bibfnamefont {V.}~\bibnamefont
  {{Cardoso}}},\ }\bibfield  {title} {\bibinfo {title} {{Ringdown of a
  dynamical spacetime}},\ }\href {https://doi.org/10.48550/arXiv.2312.04633}
  {\bibfield  {journal} {\bibinfo  {journal} {arXiv e-prints}\ ,\ \bibinfo
  {eid} {arXiv:2312.04633}} (\bibinfo {year} {2023}{\natexlab{b}})},\ \Eprint
  {https://arxiv.org/abs/2312.04633} {arXiv:2312.04633 [gr-qc]} \BibitemShut
  {NoStop}%
\bibitem [{\citenamefont {Stein}(2019)}]{Stein_qnm_2019}%
  \BibitemOpen
  \bibfield  {author} {\bibinfo {author} {\bibfnamefont {L.~C.}\ \bibnamefont
  {Stein}},\ }\bibfield  {title} {\bibinfo {title} {{qnm: A Python package for
  calculating Kerr quasinormal modes, separation constants, and
  spherical-spheroidal mixing coefficients}},\ }\href
  {https://doi.org/10.21105/joss.01683} {\bibfield  {journal} {\bibinfo
  {journal} {J. Open Source Softw.}\ }\textbf {\bibinfo {volume} {4}},\
  \bibinfo {pages} {1683} (\bibinfo {year} {2019})},\ \Eprint
  {https://arxiv.org/abs/1908.10377} {arXiv:1908.10377 [gr-qc]} \BibitemShut
  {NoStop}%
\bibitem [{\citenamefont {Leaver}(1986)}]{Leaver_1986}%
  \BibitemOpen
  \bibfield  {author} {\bibinfo {author} {\bibfnamefont {E.~W.}\ \bibnamefont
  {Leaver}},\ }\bibfield  {title} {\bibinfo {title} {Spectral decomposition of
  the perturbation response of the schwarzschild geometry},\ }\href
  {https://doi.org/10.1103/PhysRevD.34.384} {\bibfield  {journal} {\bibinfo
  {journal} {Phys. Rev. D}\ }\textbf {\bibinfo {volume} {34}},\ \bibinfo
  {pages} {384} (\bibinfo {year} {1986})}\BibitemShut {NoStop}%
\bibitem [{\citenamefont {{Carullo}}\ and\ \citenamefont {{De
  Amicis}}(2023)}]{Carullo2023}%
  \BibitemOpen
  \bibfield  {author} {\bibinfo {author} {\bibfnamefont {G.}~\bibnamefont
  {{Carullo}}}\ and\ \bibinfo {author} {\bibfnamefont {M.}~\bibnamefont {{De
  Amicis}}},\ }\bibfield  {title} {\bibinfo {title} {{Late-time tails in
  nonlinear evolutions of merging black hole binaries}},\ }\href
  {https://doi.org/10.48550/arXiv.2310.12968} {\bibfield  {journal} {\bibinfo
  {journal} {arXiv e-prints}\ ,\ \bibinfo {eid} {arXiv:2310.12968}} (\bibinfo
  {year} {2023})},\ \Eprint {https://arxiv.org/abs/2310.12968}
  {arXiv:2310.12968 [gr-qc]} \BibitemShut {NoStop}%
\bibitem [{\citenamefont {{Takahashi}}\ and\ \citenamefont
  {{Motohashi}}(2023)}]{Takahashi_2023}%
  \BibitemOpen
  \bibfield  {author} {\bibinfo {author} {\bibfnamefont {K.}~\bibnamefont
  {{Takahashi}}}\ and\ \bibinfo {author} {\bibfnamefont {H.}~\bibnamefont
  {{Motohashi}}},\ }\bibfield  {title} {\bibinfo {title} {{Iterative extraction
  of overtones from black hole ringdown}},\ }\href
  {https://doi.org/10.48550/arXiv.2311.12762} {\bibfield  {journal} {\bibinfo
  {journal} {arXiv e-prints}\ ,\ \bibinfo {eid} {arXiv:2311.12762}} (\bibinfo
  {year} {2023})},\ \Eprint {https://arxiv.org/abs/2311.12762}
  {arXiv:2311.12762 [gr-qc]} \BibitemShut {NoStop}%
\bibitem [{\citenamefont {{Qiu}}\ \emph {et~al.}(2023)\citenamefont {{Qiu}},
  \citenamefont {{Jim{\'e}nez Forteza}},\ and\ \citenamefont
  {{Mourier}}}]{Qiu2023}%
  \BibitemOpen
  \bibfield  {author} {\bibinfo {author} {\bibfnamefont {Y.}~\bibnamefont
  {{Qiu}}}, \bibinfo {author} {\bibfnamefont {X.}~\bibnamefont {{Jim{\'e}nez
  Forteza}}},\ and\ \bibinfo {author} {\bibfnamefont {P.}~\bibnamefont
  {{Mourier}}},\ }\bibfield  {title} {\bibinfo {title} {{Linear vs. nonlinear
  modelling of black hole ringdowns}},\ }\href
  {https://doi.org/10.48550/arXiv.2312.15904} {\bibfield  {journal} {\bibinfo
  {journal} {arXiv e-prints}\ ,\ \bibinfo {eid} {arXiv:2312.15904}} (\bibinfo
  {year} {2023})},\ \Eprint {https://arxiv.org/abs/2312.15904}
  {arXiv:2312.15904 [gr-qc]} \BibitemShut {NoStop}%
\end{thebibliography}%
\end{document}